\documentclass[12pt]{article}
\usepackage{cite,epsfig,amssymb,amsmath}

\usepackage[dvips]{pstricks}

\newcommand{\Eq}[1]{Eq.\,(\ref{#1})}
\newcommand{\Eqs}[1]{Eqs.\,(\ref{#1})}
\newcommand{\Eqsand}[2]{Eqs.\,(\ref{#1}) and (\ref{#2})}

\newcommand{\eq}[1]{(\ref{#1})}
\newcommand{\eqsand}[2]{(\ref{#1}) and (\ref{#2})}

\newcommand{\Fig}[1]{Figure \ref{#1}}
\newcommand{\Figsand}[2]{Figures \ref{#1} and \ref{#2}}

\newcommand{\Ref}[1]{Ref. \cite{#1}}

\newcommand{\Tab}[1]{Table \ref{#1}}

\newcommand{\Sec}[1]{Section \ref{#1}}

\newcommand{\beq}{\begin{equation}}                
\newcommand{\eeq}{\end{equation}}        
\newcommand{\bea}{\begin{eqnarray}}               
\newcommand{\eea}{\end{eqnarray}}        
\newcommand{\bdm}{\begin{displaymath}}                 
\newcommand{\edm}{\end{displaymath}}                      
  
\newcommand{\non}{\nonumber}

\newcommand{\ord}{{O}}

\newcommand{\mc}{m_c}
                
\newcommand{\mb}{m_b}   
   
\newcommand{\MT}{M_t}
\newcommand{\mll}{m_{\ell \ell}}  

\newcommand{\MW}{M_{\scriptscriptstyle W}}
\newcommand{\MZ}{M_{\scriptscriptstyle Z}}   
\newcommand{\MH}{M_{\scriptscriptstyle H}}     
  
\newcommand{\mub}{\mu_b}
  
\newcommand{\muh}{\mu_0}
\newcommand{\mut}{\mu_t}

\newcommand{\GeV}{\, {\rm GeV}}    
\newcommand{\MeV}{\, {\rm MeV}}

\newcommand{\MSbar}{ \overline{\rm MS} }

\newcommand{\gs }{ g_s }
\newcommand{\as }{ \alpha_s }         
\newcommand{\aem}{ \alpha }   
\newcommand{\f}{\frac}

\newcommand{\GF}{G_\mu}

\newcommand{\btosgamma}{b \to s \gamma}

\newcommand{\btoslpluslminus}{b \to s \ell^+ \ell^-}
\newcommand{\BtoXsgamma}{\bar{B} \to X_s \gamma}
\newcommand{\BtoXslpluslminus}{\bar{B} \to X_s \ell^+ \ell^-}

\newcommand{\BtoKKstarlpluslminus}{\bar{B} \to K^{(\ast)} \ell^+ \ell^-}

\newcommand{\BtoXclnu}{\bar{B} \to X_c \ell \nu}
\newcommand{\BtoXulnu}{\bar{B} \to X_u \ell \nu}

\newcommand{\BR}{{\rm BR}}
\newcommand{\BRll}{{{\rm BR}_{\ell \ell}}}

\newcommand\lsim{\lesssim}

\begin{document}



\thispagestyle{empty}
\rightline{CERN-TH/2003-260}
\rightline{TUM-HEP-532/03}
\rightline{FERMILAB-Pub-03/046-T}
\rightline{UCSD/PTH 03-19}
\rightline{IPPP/03/77}
\vspace*{1.3truecm}


\centerline{\LARGE \bf Complete NNLO QCD Analysis of \boldmath
$\BtoXslpluslminus$} 
\vspace{0.2cm}%
\centerline{\LARGE \bf and Higher Order Electroweak Effects}

\vskip1truecm
\centerline{\large\bf Christoph Bobeth$^{a, d}$, 
Paolo Gambino$^{b, e}$, Martin Gorbahn$^{a, f}$,}
\vspace{0.2cm}%
\centerline{\large\bf and Ulrich Haisch$^{c}$} 
\bigskip
\begin{center}{
{\em $^a$ Physik Department, Technische Universit\"at M\"unchen,
D-85748 Garching, Germany} \\ 
\vspace{.3cm}
{\em $^b$ Theory Division, CERN, CH-1211 Geneve 23, Switzerland } \\  
\vspace{.3cm}
{\em $^c$ Theoretical Physics Department, Fermilab, Batavia, IL 60510,
USA} \\
\vspace{.3cm}
{\em $^d$ Physics Department, University of California at San Diego,
La Jolla, CA 92093, USA} \\ 
\vspace{.3cm}
{\em $^e$ INFN, Sezione di Torino, 10125 Torino, Italy} \\
\vspace{.3cm}
{\em $^f$ IPPP, Physics Department, University of Durham, DH1 3LE,
Durham, UK} \\
}\end{center}
\vspace{0.75cm}

\centerline{\bf Abstract}
\vspace{0.4cm}

We complete the next-to-next-to-leading order QCD calculation of the 
branching ratio for $\BtoXslpluslminus$ including recent results for
the three-loop anomalous dimension matrix and two-loop matrix
elements. These new contributions  modify the branching ratio in the
low-$q^2$ region, $\BRll$, by about $+1 \%$ and $-4 \%$,
respectively. We furthermore discuss the appropriate normalization of
the electromagnetic coupling $\aem$ and calculate the dominant higher
order electroweak effects, showing that, due to accidental
cancellations, they change $\BRll$ by only $-1.5 \%$ if $\aem (\mu)$
is normalized at $\mu =\ord(\mb)$, while they shift it by about $-8.5 \%$ if
one uses a high scale normalization $\mu = \ord(\MW)$. The position
of the zero of the forward-backward asymmetry, $q_0^2$, is changed by
around $+2 \%$. After introducing a few additional improvements in
order to reduce the theoretical error, we perform a comprehensive
study of the uncertainty. We obtain $\BRll(1 \GeV^2 \le q^2 \le 6
\GeV^2) = (1.57 \pm 0.16) \times 10^{-6}$ and $q^2_0= ( 3.76 \pm
0.33) \GeV^2$ and note that the part of the uncertainty due to the
$b$-quark mass can be easily reduced.  

\newpage


\section{Introduction}
The rare semileptonic $\btoslpluslminus$ transitions have been
recently observed for the first time by Belle and BaBar in form of the
exclusive $\BtoKKstarlpluslminus$ modes \cite{exp}, and also inclusive
measurements are now available \cite{incl}. Like all other
flavor-changing-neutral-current transitions, these channels are
important probes of short-distance physics. Their 
study can yield useful complementary information, when confronted with the
less rare $\btosgamma$ decays, in testing the flavor sector of
the Standard Model (SM). In particular, a precise measurement of the
inclusive  $\BtoXslpluslminus$ channel would be welcome in view of
new  physics searches, because it is amenable to a clean theoretical 
description, especially for dilepton invariant masses, $\mll^2 \equiv
q^2$, below the charm resonances, namely in the range $1 \GeV^2 \lsim
\mll^2 \lsim 6 \GeV^2$. In theoretical calculations the reference
window is usually  $0.05 \le {\hat s} \equiv \mll^2/\mb^2 \le 0.25$.

The SM calculation of the Branching Ratio for $\BtoXslpluslminus$ in
this low-$\hat s$ region, $\BRll$ in the following, has reached a high
degree of maturity \cite{Bobeth:1999mk, extraNNLO, ghinculov_last,ali}: it
now includes a nearly complete resummation of large QCD logarithms $L
\equiv \ln \mb/\MW$ up to Next-to-Next-to-Leading Order (NNLO). The
peculiarity of the $\btoslpluslminus$ amplitude is that it involves
large logarithms even in the absence of QCD interactions. Factoring
out $G_\mu\aem$, it receives contributions of $\ord(\as^n L^{n +
1})$ at Leading Order (LO), 
of $\ord( \as^n L^n)$ at Next-to-Leading Order (NLO), and  of
$\ord(\as^n L^{n-1})$ at NNLO. Since the QCD penguin operators $Q_3$--$Q_6$
are suppressed by small Wilson coefficients, which makes the
computation of their two-loop matrix elements dispensable\footnote{In
$\BtoXsgamma$ the two-loop matrix elements of $Q_3$--$Q_6$ reduce the
branching ratio by around $1 \%$ \cite{Buras:2002tp}.}, the only
potentially relevant NNLO terms still missing at low $\hat s$
originate from the three-loop Anomalous Dimension Matrix (ADM) of the
operators in the low-energy effective Hamiltonian, and from the
two-loop matrix element of one of them, namely the vector-like
semileptonic operator $Q_9$. 
 Finally we would also like to remark that the NNLO matrix elements of the
current-current operators $Q_1$ and $Q_2$ for the interesting
high-$\hat s$ tail of the $\BtoXslpluslminus$ decay distributions have
been calculated just recently \cite{ginonew}.   

On the other hand, electroweak effects in $\btoslpluslminus$ have
never been studied in the literature. As shown in the case of
radiative decays \cite{ewnoi, Baranowski:1999tq, ewbsg}, they may be
as important as the higher order QCD effects. $\BRll$ is generally
parameterized in terms of the electromagnetic coupling $\aem$, but the
scale at which $\aem$ should be evaluated is, in principle,
undetermined until higher order electroweak effects are taken into
account. This has led most authors to use $\aem (\mll) \approx \aem
(\mb) \approx 1/133$. Indeed, as we will discuss later on, in the
absence of an $\ord (\aem)$ calculation, there is no reason to 
consider $\aem(\mll)$ more appropriate than, say, $\aem (\MW) \approx
1/128$. As $\BRll$ is proportional to $\aem^2$, the ensuing
uncertainty  of almost $8 \%$ is not at all negligible.  

The purpose of this paper is threefold: $i)$ in \Sec{sec:NNLO} we
evaluate the two main missing NNLO QCD contributions to
$\BtoXslpluslminus$ taking advantage of the new results for the
three-loop ADM \cite{Gambino:2003zm, next} and of existing
calculations of $\ord (\as^2)$ corrections to semileptonic quark
decays; $ii)$ in \Sec{sec:EW} we study to what extent higher order
electroweak effects influence this decay mode by calculating the
dominant $\ord (\aem)$ contributions to the running of the Wilson
coefficients and estimating other potentially large effects; $iii)$ in
\Sec{sec:PH} we update the SM prediction of $\BRll$ and  of the
position of the zero of
the Forward-Backward (FB) asymmetry, $\hat{s}_0$, using the above 
results, and discuss a few improvements aimed at reducing the
residual theoretical error in those observables.      

\section{Completing the NNLO QCD Calculation}
\label{sec:NNLO}

Let us first focus on the missing NNLO QCD contributions. The main
steps of the NNLO calculation have been outlined in
\cite{Bobeth:1999mk}, while two-loop matrix elements and
bremsstrahlungs corrections have been computed in 
\cite{extraNNLO}. 
In the following we adopt the 
operator basis of  \cite{Bobeth:1999mk}, which we enlarge to include
the electroweak penguin operators $Q^Q_3$--$Q^Q_6$ and the operator
$Q^b_1$, in view of our subsequent discussion of electroweak effects:
\begin{alignat}{3} \label{list}
Q_1 & = (\bar{s}_L \gamma_\mu T^a c_L) (\bar{c}_L \gamma^\mu T^a b_L) 
\, , & Q_2 & = (\bar{s}_L \gamma_\mu c_L) (\bar{c}_L \gamma^\mu b_L)
\, , \non \\[1mm]    
Q_3 & = (\bar{s}_L \gamma_\mu b_L) \scalebox{1.0}{$\sum\nolimits_q$}
(\bar{q} \gamma^\mu q) \, , & Q_4 & = (\bar{s}_L \gamma_\mu T^a b_L)
 \scalebox{1.0}{$\sum\nolimits_q$} (\bar{q} \gamma^\mu T^a q) \, , \non
\\[1mm]    
Q_5 & = (\bar{s}_L \gamma_\mu \gamma_\nu \gamma_\rho b_L)
 \scalebox{1.0}{$\sum\nolimits_q$} (\bar{q} \gamma^\mu \gamma^\nu
\gamma^\rho q) \, , & 
Q_6 & = (\bar{s}_L \gamma_\mu \gamma_\nu \gamma_\rho T^a b_L)
 \scalebox{1.0}{$\sum\nolimits_q$} (\bar{q} \gamma^\mu \gamma^\nu
\gamma^\rho T^a q) \, , \non \\[1mm]
Q^Q_3 & = (\bar{s}_L \gamma_\mu b_L)  \scalebox{1.0}{$\sum\nolimits_q$}
Q_q (\bar{q} \gamma^\mu q) \, , & Q^Q_4 & = (\bar{s}_L \gamma_\mu T^a
b_L) \scalebox{1.0}{$\sum\nolimits_q$} Q_q (\bar{q} \gamma^\mu T^a q) \,
, \non \\[1mm]     
Q^Q_5 & = (\bar{s}_L \gamma_\mu \gamma_\nu \gamma_\rho b_L)
 \scalebox{1.0}{$\sum\nolimits_q$} Q_q (\bar{q} \gamma^\mu \gamma^\nu
\gamma^\rho q) \, , & Q^Q_6 & = (\bar{s}_L \gamma_\mu \gamma_\nu
\gamma_\rho T^a b_L) \scalebox{1.0}{$\sum\nolimits_q$} Q_q (\bar{q}
\gamma^\mu \gamma^\nu \gamma^\rho T^a q) \, , \non \\[1mm]
Q^b_1 & = - \scalebox{1.0}{$\f{1}{3}$} (\bar{s}_L \gamma_\mu b_L)
(\bar{b} \gamma^\mu b) + \scalebox{1.0}{$\f{1}{12}$} (\bar{s}_L
\gamma_\mu \gamma_\nu \gamma_\rho b_L) (\bar{b} \gamma^\mu \gamma^\nu
\gamma^\rho b) \, , \hspace{-2.75cm} & & \non \\[1mm]  
Q_7 & = \scalebox{1.0}{$\f{e}{\gs^2}$} \mb (\bar{s}_L \sigma^{\mu \nu}
b_R) F_{\mu \nu} \, , & Q_8 & =  \scalebox{1.0}{$\f{1}{\gs}$} \mb
(\bar{s}_L \sigma^{\mu \nu} T^a b_R) G_{\mu \nu}^a \, , \non \\[0.5mm]
Q_9 & = \scalebox{1.0}{$\f{e^2}{\gs^2}$} (\bar{s}_L \gamma_\mu b_L)
\scalebox{1.0}{$\sum\nolimits_\ell$} (\bar{\ell} \gamma^\mu \ell) \, , &
Q_{10} & = \scalebox{1.0}{$\f{e^2}{\gs^2}$} (\bar{s}_L \gamma_\mu b_L) 
\scalebox{1.0}{$\sum\nolimits_\ell$} (\bar{\ell} \gamma^\mu \gamma_5
\ell) \, .  
\end{alignat}
The operator $Q_1^b$ corresponds in four dimensions to
$(\bar s_L \gamma^\mu b_L)(\bar b_L \gamma_\mu b_L)$ and receives
contributions from electroweak boxes \cite{Q2b}. In order to avoid the
introduction of traces with $\gamma_5$ at all orders in QCD, we have
rewritten this operator in a slightly more involved form\footnote{We
thank M.~Misiak for this suggestion.}.

The renormalization scale dependence of the Wilson coefficients
${\vec C}^T (\mu) = ( C_1 (\mu), \ldots, \\ C_{10} (\mu))$ of the
effective operators is governed by the Renormalization Group Equation
(RGE) whose solution is schematically given by  
\beq \label{cmu}
{\vec C} (\mu) = {\hat U} (\mu,\muh,\aem) {\vec C} (\muh) \, .   
\eeq
In the case at hand we are interested in the running of the Wilson
coefficients from the electroweak scale $\muh = \ord (\MW)$ down to
the low-energy scale $\mub =\ord (\mb)$. Neglecting for the moment the
electromagnetic coupling $\aem$,  all the QCD corrections
to the initial Wilson coefficients relevant at  NNLO have explicitly
been given in \cite{Bobeth:1999mk}, while the only $\ord(\as^2)$
contributions to the evolution matrix ${\hat U} (\mu,\muh,\aem)$
pertinent to $\btoslpluslminus$ at NNLO concerns the
mixing of $Q_2$ into $Q_7$ and $Q_9$. Expanding ${\hat U}
(\mu,\muh,\aem)$ in powers of $\as (\mub)/4\pi$ --- see \Eq{utot}   
--- we denote these terms by $U_{s, 72}^{(2)}(\mub,\muh)$ and $U_{s,
92}^{(2)}(\mub,\muh)$, respectively\footnote{The definition of 
$U^{c (2)}_{92} (\mub,\muh)$ adopted in Eq.~(37) of
\Ref{Bobeth:1999mk} corresponds to $\eta U_{s, 92}^{(2)} (\mub,\muh)$ 
used here.}. The ingredients necessary for the calculation of $U_{s,
72}^{(2)}(\mub,\muh)$ were already available four years ago, and are
included in \cite{Bobeth:1999mk}. On the other hand, the calculation
of $U_{s, 92}^{(2)}(\mub,\muh)$ requires the knowledge of the
three-loop self-mixing of $Q_1$--$Q_6$ as well as the three-loop
mixing of $Q_1$--$Q_6$ into $Q_{9}$. The relevant ADM entries have
been calculated only recently in \cite{Gambino:2003zm, next}. Solving
the NNLO RGE \cite{next, beneke}, we obtain   
\beq \label{u92}
U_{s, 92}^{(2)}(\mub,\muh) = \sum_{i = 1}^9 \left[ b_i \eta^{a_i} +
c_i \eta^{a_i + 1} + d_i \eta^{a_i + 2} \right] \approx  181.42 \, 
e^{-5.1901 \, \eta} \, \eta^{1.5212} \, ,  
\eeq
where $\eta = \as(\muh)/\as(\mub)$, and the so-called magic numbers
$a_i$, $b_i$, $c_i$ and $d_i$ can be found in \Tab{tab:magic}. Using
$\as (\MZ) = 0.119$ and $\mb = 4.8 \GeV$, we find numerically
$U_{s, 92}^{(2)}(\mb, \MW) \approx 4.1$. Unless $\eta < 0.48$, our 
result is within the range that was guessed in \cite{Bobeth:1999mk},
$-10 \, \eta \le   U_{s, 92}^{(2)}(\mub,\muh) \le 10 \, \eta$. Our
determination of $U_{s, 92}^{(2)}(\mub,\muh)$ eliminates one source of
uncertainty in the NNLO calculation --- we will study the residual
scale dependence later on --- and increases the value of $\BRll$
presented in \Sec{sec:PH} by about $1 \%$, the exact amount depending
on the choice of the various renormalization scales.    

{%
\renewcommand{\arraycolsep}{3.5pt}
\begin{table}[t] 
\begin{center}
\[
\begin{array}{|c|ccccccccc|}\hline
i & 1 & 2 & 3 & 4 & 5 & 6 & 7 & 8 & 9 \\
\hline a_i & \scalebox{0.9}{$\f{6}{23}$} & -\scalebox{0.9}{$\f{12}{23}$} &
0.4086 & -0.4230 & -0.8994 & 0.1456 & -1 & -\scalebox{0.9}{$\f{24}{23}$}
& \scalebox{0.9}{$\f{3}{23}$} \\[0.5mm] 
b_i & 12.4592 & 0.6940 & -1.7339 & 1.2360 & -0.1921 & 0.3998 & 0 & 0 &
0 \\[0.5mm] 
c_i & -2.5918 & -0.2971 & -0.5949 & 0.1241 & 0.3170 & 2.8655 & 0 & 0 &
0 \\[0.5mm] 
d_i & 1.3210 & 3.1616 & -0.4814 & 1.9362 & -5.0873 & 0.0468 & -13.5825
& 0 & 0 \\ 
\hline e_i & -0.0238 & 0.0107 & 0.0023 & 0.0071 & 0.0050 & -0.00003 &
-0.0087 & -0.0008 & 0.0342 \\[0.5mm] 
f_i & 0.0010 & 0.0130 & 0.0045 & -0.0022 & -0.0714 & -0.0008 & 0.0299
& 0 & 0 \\  
\hline g_i & 0 & 0 & 0 & 0 & 0 & 0 & 0.0035 & 0 & 0 \\[0.5mm] 
h_i & 0.0114 & -0.0107 & -0.0012 & -0.0057 & -0.0098 & 0.0002 & 0.0122
& 0 & 0 \\ 
\hline
\end{array}
\]
\end{center}
\caption{\sf Numerical coefficients parameterizing the RGE solutions
in \Eqs{u92}, \eq{ue92} and \eq{ue102}.}
\label{tab:magic}
\end{table} 
}%

\begin{figure}[t]
\hspace{2.5cm}
\scalebox{0.5}{\includegraphics{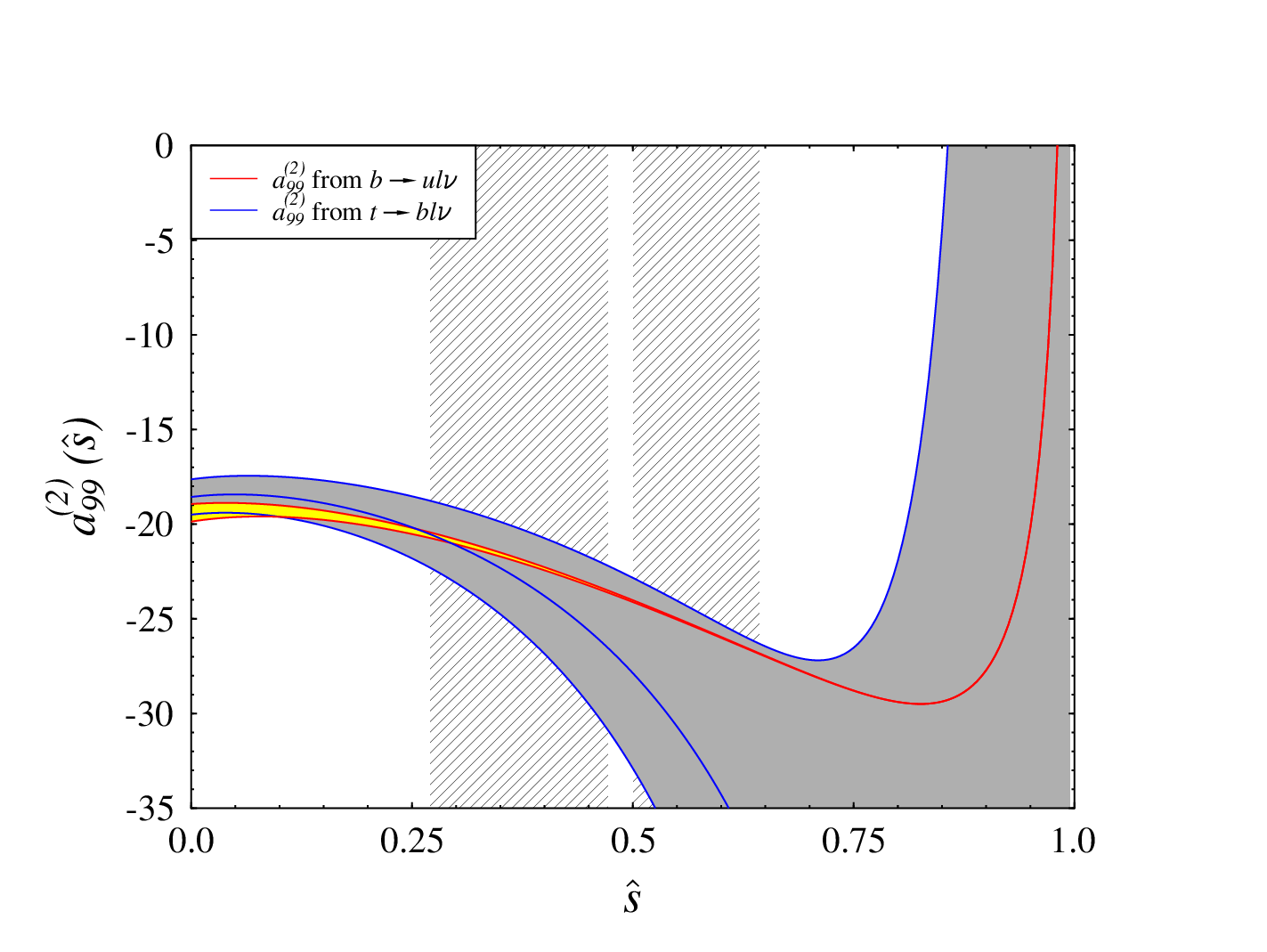}}

\vspace{-2mm}
\caption{\sf Second order perturbative corrections to the 
matrix element of $Q_9$, see \Eq{me9}, as a function of the 
dilepton invariant mass.
The lower (upper) red curve corresponds to
the $1 - \hat s$ expansion up to third (fourth) order \cite{btou},
while the blue curves correspond to the $\hat s$ expansion 
\cite{tdecay} (central value and linearly added errors). The shaded  
areas show the cuts of the experimental analysis of the Belle and
BaBar Collaborations \cite{incl} to reject backgrounds due to $J/\psi$
resonances. 
\label{fig:ME}} 
\end{figure}

Another missing ingredient of the complete NNLO calculation is the
two-loop $\ord(\as^2)$   
matrix element of $Q_9$. This contribution is necessary because $Q_9$
has a non-vanishing matrix element at tree-level as well as a non-zero
Wilson coefficient in leading logarithmic approximation.  
Fortunately, no explicit calculation is necessary in this case as the
QCD corrections to $\btoslpluslminus$ are identical to those to $b \to
u \ell \nu$ ($t \to b \ell \nu$), in the limit of vanishing strange
(bottom) quark  mass. In particular, we want to have the QCD
corrections to the dilepton invariant mass spectrum. The $\ord
(\as^2)$ corrections to  this spectrum for the decay $b\to u \ell \nu$
have been computed in \cite{btou} in terms of an expansion in $1 -
{\hat s}$, which converges also well in the low-${\hat s}$ region of
interest. The results up to third and fourth order in the $1 - {\hat
s}$ expansion are shown in \Fig{fig:ME},  
in the $b$-quark pole mass scheme.

The dilepton invariant mass spectrum at small $\hat s$ can also be obtained
from the $\MW^2/\MT^2$ expansion of the second order QCD corrections
to the top quark decay calculated in \cite{tdecay}. In principle, using the
replacements $\MW \to \mll$ and $\MT \to \mb$, this expansion is
better suited for the low-$\hat s$ region. However, \Ref{tdecay} 
provides only the terms up to $\MW^4/\MT^4$, obtained through a Pad\'e
approximation from a $q^2/\MT^2$ expansion. The ensuing uncertainty
is displayed in \Fig{fig:ME}, where the errors of the various
coefficients have been added linearly and as before the pole mass 
scheme for the $b$-quark mass has been used. Although this is likely
to overestimate the uncertainty in the calculation based on
\cite{tdecay}, the precision achieved is sufficient for our
purposes. Moreover, \Fig{fig:ME} shows that the two  approaches
\cite{btou, tdecay} to the calculation of the $\ord (\as^2)$
corrections to the invariant dilepton mass spectrum agree quite
well in the low-$\hat s$ region. Normalizing the  matrix element to
its tree level value,   
\beq\label{me9}
\big | \left \langle Q_9 \right \rangle \big |^2 = \big | \left
\langle Q_9 \right \rangle^{(0)} \big |^2 \left [ 1 + \f{\as
(\mb)}{\pi} a_{99}^{(1)} (\hat s) + \left(\f{\as (\mb)}{\pi} 
\right)^2 a_{99}^{(2)} (\hat s) \right] \, , 
\eeq
a simple approximation of the real and virtual corrections denoted 
by $a_{99}^{(2)} (\hat s)$ is 
\beq
  a_{99}^{(2)} (\hat s) \approx 
  \frac{-19.2 + 6.1 \, \hat s + \left ( 37.9 + 17.2 \ln \hat s \right ) \hat s^2 
   - 18.7 \, \hat s^3 }{(1+ 2\hat s) \left ( 1-\hat s\right)^2}  \, .
\eeq
This approximate formula is valid  
in the range $0 \le \hat s \le 0.4$, in the pole mass 
scheme for the $b$-quark mass. Using this expression in the NNLO
calculation of $\BRll$ presented in \Sec{sec:PH}, we observe a 
reduction of about $4 \%$, that overcompensates the three-loop
running of $C_9 (\mub)$ discussed above.

\section{Higher Order Electroweak Effects}
\label{sec:EW}

\begin{figure}[t]
\begin{center}
\scalebox{0.8}{\begin{picture}(75,45)(0,0)\includegraphics{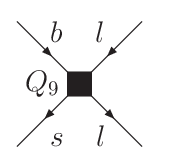}\end{picture}}
\hspace{2mm}
\scalebox{0.8}{\begin{picture}(120,45)(0,0)\includegraphics{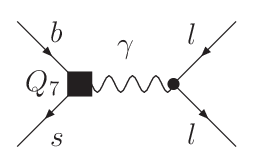}\end{picture}}
\hspace{2mm}
\scalebox{0.8}{\begin{picture}(140,45)(0,2)\includegraphics{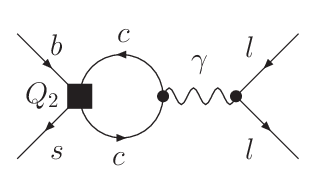}\end{picture}}
\end{center}
\vspace{-.35cm}
\caption{\sf Some of the basic diagrams contributing to
$\btoslpluslminus$ in the effective theory.} 
\label{fig:LO}
\end{figure}

A comprehensive study of electroweak effects has been performed in the
case of $\BtoXsgamma$ \cite{ewnoi, Baranowski:1999tq, ewbsg} but, to
the best of our knowledge, such corrections have never been considered
in the calculation of $\BRll$. Unlike in the $\btosgamma$ processes,
only virtual photons appear in the basic $\btoslpluslminus$
transition. In the following, we will try to identify and estimate
the dominant $\ord (\aem)$ electroweak effects. When we consider large
logarithms we refer only to terms like $\ln \mb/\MW$ and exclude the
case of very small ${\hat s}$, which would also involve large logarithms
of the form $\ln \hat s$.   

Let us first concentrate on the origin of large logarithms at
$\ord(\aem)$. To this end, it is best to work from the very
beginning in the effective theory at lowest order in the
electromagnetic coupling. At the weak scale, we distinguish between
two kinds of contributions to the $\btoslpluslminus$ amplitude: $a)$
those containing a virtual photon and $b)$ those that do not include  
photons at all. Examples of the first kind are $Q_2$ and $Q_7$
insertions, while $Q_9$ and $Q_{10}$ insertions belong to the second
category --- see \Fig{fig:LO}. This classification holds true also at
higher orders in QCD.  

\begin{figure}[t]
\begin{center}
\scalebox{0.8}{\begin{picture}(160,45)(0,0)\includegraphics{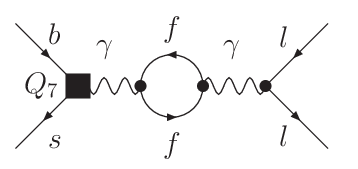}\end{picture}}
\hspace{2mm}
\scalebox{0.8}{\begin{picture}(160,45)(0,-1.25)\includegraphics{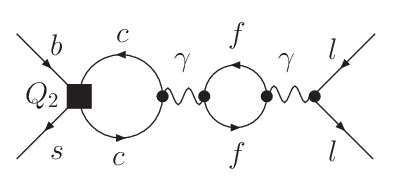}\end{picture}}
\end{center}
\vspace{-.35cm}
\caption{\sf Vacuum polarization diagrams.} 
\label{fig:VP}
\end{figure}

\begin{figure}[t]
\begin{center}
\scalebox{0.8}{\begin{picture}(90,45)(0,0)\includegraphics{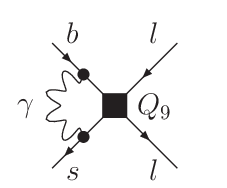}\end{picture}}
\hspace{2mm}
\scalebox{0.8}{\begin{picture}(140,45)(0,0)\includegraphics{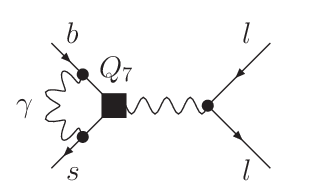}\end{picture}}
\hspace{2mm}
\scalebox{0.8}{\begin{picture}(160,45)(0,15)\includegraphics{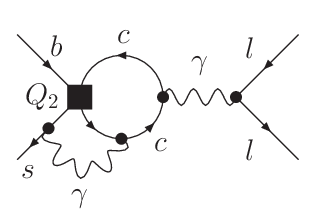}\end{picture}}
\end{center}
\vspace{-.35cm}
\caption{\sf Some diagrams contributing to the running
of the Wilson coefficients.}
\label{fig:NLO}
\end{figure}

Evolving from higher to lower scales, the contributions of class $a)$ 
receive $\ord (\aem)$ corrections involving large logarithms related
either to the running of the QED coupling constant or to the evolution
of the relevant Wilson coefficients. The first effect has its origin
in vacuum polarization diagrams on the virtual photon propagator
proportional to 
$\aem \ln \mll/\mu$, like the one shown in \Fig{fig:VP}, that
can be effectively taken into account by changing the scale at which
$\aem$ is evaluated in the overall factor of $\BRll$ from  the
weak scale to $\mll$.
The second effect has to do with the fact that, in full
analogy to what happens in QCD, photonic interactions induce an ADM
for the effective operators --- see \Fig{fig:NLO}. In order to include
all $\ord (\aem)$ logarithms, it is therefore sufficient to evolve
both the photonic coupling and the Wilson coefficients according to a
mixed QED--QCD RGE from the matching scale $\muh = \ord(\MW)$ to a
low-energy scale $\mu = \ord(\mb)$. Beyond leading order in $\aem$,
one can simply normalize the amplitudes of type $a)$ with $\GF \aem
(\mu)$, and evolve the Wilson coefficients at $\aem$ fixed, as done in
\cite{ewnoi, ewbsg}. 

Concerning contributions of class $b)$, no vacuum polarization diagram
can contribute at $\ord (\aem)$, and the only $\ord (\aem)$ large
logarithms are related to the running of the operators. As the
normalization of the coupling constant is determined by the short
distance interactions at the matching scale, the  Wilson coefficients 
are appropriately expressed in terms of $\GF$ using the relation 
$g^2 (\MW)/4\pi = \aem (\MW)/\sin^2
\theta_{\scriptscriptstyle W} (\MW) = \sqrt{2}/\pi \, \GF \MW^2$,
 where all running couplings are in the $\MSbar$
scheme. The typical case is that of box diagrams, which are clearly
proportional to  $g^4 (\MW)/\MW^2 \propto \GF^2 \MW^2$. After
decoupling, all short distance information is encoded in the
Wilson coefficients and in $\GF$, which does {\it not} evolve in the
effective theory.  The coupling $e^2$ in front of $Q_9$ and $Q_{10}$ --- see  
\Eq{list} --- is naturally evaluated at the weak scale.

All electroweak couplings
entering the matching can be expressed in terms of $G_\mu$, $\MW$, and
$\sin^2\theta_{\scriptscriptstyle W} (\MW)$.  These parameters are 
frozen in the low-energy effective theory. On the other hand, the
electromagnetic coupling of the photon in the low-energy effective
theory does run and the low-energy matrix elements must be expressed
in terms of the low-energy  gauge couplings $\as(\mu)$ and
$\aem(\mu)$, and of the Wilson coefficients $\vec C(\mu)$.  
This simple intuitive picture is naturally realized by a mixed QED-QCD 
RGE.  In fact, a general feature
of the Renormalization Group (RG) evolution is to keep the 
coupling in front of the Wilson coefficients fixed at the high
scale via factors of $\aem(\muh)/\aem(\mu)$. 

The only complication we face here is due to our choice of the
operator basis. The factor $e^2/g_s^2$ in front of the operators
$Q_9$ and $Q_{10}$  is such that  the Wilson coefficients at the high
scale $C_9(\muh)$ and $C_{10}(\muh)$ start at $O(\aem_s G_\mu)$ rather
than  at $O(\aem G_\mu)$.  As a consequence, the evolution of the
coefficients  $C_9 (\mu)$ and $C_{10} (\mu)$ does not {\it explicitly}
include a  rescaling factor $\aem(\muh)/\aem(\mu)$ 
as, for instance, in the standard QCD evolution of  $C_1 (\mu)$
and $C_2 (\mu)$. 
However, we will show at the end of this section that 
the correct normalization of the electroweak 
coupling in front of these contributions is preserved by the 
QED $\beta$ function coefficients 
$\beta_e^{(0)}$ and $\beta_{es}^{(1)}$ that appear in 
the self-mixing of $Q_{9}$ and $Q_{10}$ --- see
\Eqsand{eq:gammae0}{eq:gammae1} --- due to the  $e^2$ factor in the
definition of these operators.
 
Summarizing, we perform the RG evolution of 
the Wilson coefficients which
results in the normalization of the whole amplitude in terms of
$\aem(\mu)$. As we
will discuss later on, additional sub-dominant corrections come from 
$\ord (\aem)$ matrix elements of the effective operators and from
$\ord (\aem)$ corrections to the matching conditions.   
We employ  $\sin^2
\theta_{\scriptscriptstyle W} \equiv \sin^2\hat
\theta_{\scriptscriptstyle W}^{\MSbar} (\MZ) = 0.2312$, $\MT = \MT 
(\MT) = 165 \GeV$ and $\MW = 80.426 \GeV$. We adopt an $\MSbar$ 
definition of $\aem (\mu)$ with $\aem^{(5)}(\MZ) = 1/127.765$
\cite{alphamz} as input.  


Let us now study  explicitly the QED--QCD evolution of the Wilson
coefficients. Our basis \eq{list} differs from  the one used in
\cite{ewnoi} by a factor $\gs^{-2}$ in the normalization of
$Q_7$--$Q_{10}$, which complicates slightly the counting of couplings
and the comparison with the latter paper. Neglecting the running of
$\aem$ --- an $\ord (\aem^2)$ effect in the running of the
Wilson coefficients --- we can expand the evolution matrix ${\hat U}
(\mu,\muh,\aem)$ of \Eq{cmu} in the following way  
\beq \label{utot}
\begin{split}
{\hat U} (\mu,\muh,\aem) & = {\hat U}^{(0)}_s (\mu, \muh) + 
\f{\as(\mu)}{4 \pi} {\hat U}^{(1)}_s (\mu, \muh) + \left (
\f{\as(\mu)}{4 \pi}\right )^2 {\hat U}^{(2)}_s (\mu, \muh) \\ 
& + \f{\aem}{\as(\mu)} {\hat U}_e^{(0)} (\mu, \muh) + \f{\aem}{4\pi}
\hat U_e^{(1)} (\mu, \muh) +  \f{\aem \as (\mu)}{(4\pi)^2} \hat
U_e^{(2)} (\mu, \muh) + \ldots \, .
\end{split}
\eeq
The matrices $\hat U^{(i)}_s (\mu, \muh)$ and $\hat U_e^{(i)} (\mu,
\muh)$ describe pure QCD and mixed QED--QCD evolution, respectively, and
are functions of the ADM\footnote{\Eq{ADM} contains also a term
proportional to $\aem^2/\as (\mu)$, from the mixing of $Q_9$ into
$Q^Q_3$. We have checked that its contribution to the evolution of the
relevant Wilson coefficients is negligibly small.} 
\beq \label{ADM}
\begin{split}
\hat{\gamma} (\mu) & = 
\f{\as (\mu)}{4 \pi} {\hat \gamma}_s^{(0)} + \left( \f{\as (\mu)}{4
\pi} \right)^2 {\hat \gamma}_s^{(1)} + \left( \f{\as (\mu)}{4
\pi} \right)^3 {\hat \gamma}_s^{(2)} \\ 
& + \f{\aem}{4 \pi} {\hat \gamma}_e^{(0)} + \f{\aem \as (\mu)}{(4
\pi)^2} {\hat \gamma}_e^{(1)} + \dots \, ,  
\end{split}
\eeq
of the operators in question and of the QCD, QED and mixed QED--QCD $\beta$
functions.
The matrices
$\hat U_e^{(0)} (\mu, \muh)$ and $\hat U_e^{(1)} (\mu,\muh)$ 
can be computed using the formalism described for instance in
\cite{Buras:1993dy}. Expanding also the Wilson coefficients at the
weak scale  
\beq \label{cfin}
\begin{split}
\vec C (\muh) & = \vec C^{(0)}_s (\muh) + \f{\as (\muh)}{4 \pi} \vec 
C^{(1)}_s (\muh) + \left ( \f{\as (\muh)}{4 \pi} \right )^2 \vec
C^{(2)}_s (\muh) \\ 
& + \f{\aem}{4 \pi} \vec C_e^{(1)} (\muh) + \f{\aem
\as(\muh)}{(4\pi)^2} \vec C_{e}^{(2)} (\muh) + \ldots \, ,   
\end{split}
\eeq
and inserting \eqsand{utot}{cfin} into \eq{cmu}, we find the
expressions for the various terms at the low scale $\mu$ 
\beq
\begin{split}
\vec C (\mu) & = \vec C^{(0)}_s (\mu) + \f{\as (\mu)}{4 \pi} \vec 
C^{(1)}_s (\mu) + \left ( \f{\as (\mu)}{4 \pi} \right )^2 \vec
C^{(2)}_s (\mu) \\ 
& + \f{\aem}{\as (\mu)} \vec C_e^{(0)} (\mu) + \f{\aem}{4 \pi} \vec
C_e^{(1)} (\mu) + \f{\aem \as(\mu)}{(4 \pi)^2} \vec C_{e}^{(2)} (\mu)
+ \ldots \, .  
\end{split}
\eeq
Here the coefficients $\vec C^{(i)}_s (\mu)$ result from
the  $\ord (1)$, $\ord (\as)$ and $\ord (\as^2)$ contributions to
$\vec C (\muh)$ and from the QCD evolution matrices $\hat U^{(i)}_s
(\mu,\muh)$, whereas  the $\vec C^{(i)}_e (\mu)$ stem
from the various $\ord (\aem)$ corrections to $\vec C (\muh)$ and the
QCD, QED--QCD evolution matrices $\hat U^{(i)}_s (\mu,\muh)$ and $\hat
U^{(i)}_e (\mu,\muh)$.     

The formally leading electroweak effects are $\ord(\aem \as^{n - 1}
L^{n + 1})$ in the amplitude 
and are contained in 
\beq 
\vec C_e^{(0)} (\mu) = {\hat U}_e^{(0)} (\mu, \muh) {\vec C}_s^{(0)}
(\muh) \, . 
\eeq 
The non-vanishing $\ord (\aem)$ mixing described by $\hat
\gamma_e^{(0)}$ has been calculated in \cite{Baranowski:1999tq},
except for the QED mixing of $Q_9$ and $Q_{10}$. The lowest order ---
$\ord (\as)$ in our notation --- mixing between the electroweak
penguin operators $Q^Q_3$--$Q^Q_6$ and $Q_{9}$ is also missing in the 
literature, as is the $\ord (\as)$ mixing of the operator
$Q^b_1$. We have computed all missing entries and report the full
matrices $\hat \gamma_e^{(0)}$ and $\hat \gamma_s^{(0)}$ in the
Appendix. Since only the mixing of $Q_2$ into $Q_{9}$ and $Q_{10}$ is 
relevant at this order, we solve the RGE and get   
\beq \label{ue92}
C_{e, 9}^{(0)} (\mu) = \sum_{i = 1}^{9} \left[ e_i \eta^{a_i - 1} + f_i
\eta^{a_i} \right] \approx 5.7947 \, e^{-11.508 \, \eta} \,
\eta^{2.8808} \, , 
\eeq
where we used $C_2^{(0)} (\muh) = 1$, and the magic numbers $a_i$, 
$e_i$ and $f_i$ are given in \Tab{tab:magic}. The approximation is
valid within less than $1 \%$ for $0.5 \le \eta \le 0.6$. 
The formally leading QED contribution shifts $C_9 (\mb)$ by only
$0.00006$ for $\eta = 0.56$. 
 Since $Q_9$ mixes into $Q_{10}$ at $\ord (\aem)$ we also
have a contribution to
\beq \label{ue102}
C_{e,10}^{(0)} (\mu) = \sum_{i = 1}^{9} \left[ g_i \eta^{a_i - 1} +
h_i \eta^{a_i} \right] \approx 1.9013 \, e^{-9.7876 \, \eta} \,
\eta^{1.2089} \, ,  
\eeq
which shifts $C_{10} (\mb)$ by $0.00014$ for $\eta = 0.56$. Again the
numerical coefficients $a_i$, $g_i$ and $h_i$ can be found in
\Tab{tab:magic}, and the approximation is valid within less than $1
\%$ for $0.5 \le \eta \le 0.6$. Since in our normalization $C_9
(\mb) \simeq 0.072$ and $C_{10} (\mb) \simeq -0.072$ at NNLO QCD, the
impact of these $\ord (\aem/\as)$ corrections on $\BRll$ is tiny.

As discussed in \cite{ewnoi, Buras:1999st}, next-to-leading
electroweak effects can be larger than the leading ones. Moreover, in
the case of $\btoslpluslminus$, the LO QCD contribution is
accidentally small compared to the NLO QCD contribution. Let us therefore
investigate the importance of these $\ord(\aem \as^{n - 1} L^n)$
effects. The general expression for $\vec C_e^{(1)} (\mu)$ is 
\beq \label{ce1}
\vec C_e^{(1)} (\mu) = \hat U^{(0)}_s (\mu, \muh) \vec C_e^{(1)}(\muh)
+ \eta \, \hat U_e^{(0)} (\mu, \muh) \vec C_s^{(1)}(\muh) + \hat
U_e^{(1)} (\mu, \muh) \vec C_s^{(0)}(\muh) \, .  
\eeq
The last term in this equation requires the knowledge of the
$\ord (\aem \as)$ ADM $\hat \gamma^{(1)}_e$, which we have calculated
using the method described in \cite{Gambino:2003zm}. The result is
given in the Appendix. Rather than giving analytic expressions for
these $\ord (\aem)$ corrections to the Wilson coefficients of $Q_7$,
$Q_9$ and $Q_{10}$, we  only provide some approximate formulas and
refer to a forthcoming publication \cite{future} for more
details. Taking into account the different operator normalization, we 
 find the same results as in \cite{ewnoi}
for the $\ord (\aem)$ correction to the Wilson coefficient of the
electromagnetic operator,   
\beq \label{c7e1}
\begin{split}
C_{e,7}^{(1)} (\mu) & \approx 0.2894 \, e^{-5.1605 \, \eta} \, 
\eta^{1.4543} x_t^{0.3060} + 10654 \, e^{-14.857 \, \eta} \,  
\eta^{8.2193} \\[1mm]  
& \approx  0.0107 + 0.0221 \, , 
\end{split}
\eeq
where we have distinguished between the last two terms on the
right-hand side of \Eq{ce1} --- in the particular case of $C_7 (\mu)$
the first term is exactly zero as there is no $\ord (\as)$ mixing of
the four-quark operators into the magnetic one, and the $\ord (\aem)$ 
contribution to $C_7 (\muh)$ also vanishes in the normalization
employed here, see \Eq{list}. The approximate formulas given above  
and in the following are all valid within $1 \%$ for $0.5 \le \eta \le
0.6$ and $160 \GeV \le \MT \le 170 \GeV$. Furthermore, the last line
of the latter equation gives the numerical values of the
individual contributions for $\eta = 0.56$ and $\MT(\MT) = 165
\GeV$. All together, these $\ord(\aem)$ contributions lead to a
shift of $0.00002$ in $C_7 (\mb)$. Given that in our normalization
$C_7 (\mb) \simeq -0.0053$ at NNLO QCD, this  correction 
is quite small. The analogous expression for $C_9 (\mu)$ reads    
\beq \label{c9e1}
\begin{split}
C_{e,9}^{(1)} (\mu) & \approx 
-41.496\,e^{-4.8790 \,\eta} \,\eta^{0.4747}\, x_t^{0.4738}
+75.537 e^{-5.4866\,\eta}\, \eta^{0.4931}\, x_t^{- 1.9596}
 \\[1mm] 
& + 13.565 \, e^{-6.0599 \,\eta} \, \eta^{2.3409} 
\\[1mm] 
& \approx -4.0517 + 0.1579 + 0.1173 \, .   
\end{split}
\eeq
Here and in the following numerical results we have set $\muh = \MW$. 
We have
distinguished between the three terms in \Eq{ce1}. Again, the last line
in the above equation gives the numerical value of the three
contributions for $\eta = 0.56$ and $\MT(\MT) = 165 \GeV$: this electroweak
correction to $C_9 (\mb)$ is around $-0.00226$,  thus much larger than
the one in \Eq{ue92}. Moreover, it is dominated by the first term
in \Eq{ce1}, namely by the $\ord (\aem)$ matching
conditions,  and particularly by those of the electroweak
penguin operators. Numerically, it corresponds to approximately $-3\%$ of the
NNLO QCD coefficient. The corresponding expression for $C_{10}
(\mu)$ is   
\beq \label{c10e1}
\begin{split}
C_{e,10}^{(1)} (\mu) & \approx 
-20.835 \, e^{-4.9135 \,\eta} \,\eta^{0.4957} \,x_t^{0.8394}
+ 67049 \, e^{-15.347 \, \eta} \,
\eta^{8.8810} \\[1mm]    
& \approx -3.3340 + 0.0720\, .  
\end{split}
\eeq
The shift in $C_{10} (\mb) $ is about $-0.00195$, and is again
dominated by a term stemming from the higher order matching
corrections. Like in the case of $C_7 (\mu)$, the term resulting from
the leading order QCD running is exactly zero since there is no mixing
of the four-quark operators into $Q_{10}$ at $\ord (\as)$, and the
$\ord (\aem)$ correction to $C_{10} (\muh)$ is zero too. Numerically,
it corresponds to approximately $+3 \%$ of the NNLO QCD coefficient.

The experience with radiative decays \cite{ewnoi} shows that
electroweak corrections to the matching conditions may end up being
the dominant electroweak contributions to the low scale
coefficients. In the notation adopted here, such contributions to
$C_7 (\muh)$, $C_9 (\muh)$ and  $C_{10} (\muh)$ are of $\ord (\aem
\as)$ and therefore formally enter at NNLO or $\ord(\aem\as^{n}L^n)$
in the $\btoslpluslminus$ amplitude. Although a complete analysis at
this order is clearly beyond the scope of our paper, it is worth
investigating the magnitude of the potentially dominant terms, such as
electroweak effects in the matching enhanced  by factors $1/\sin^2 
\theta_{\scriptscriptstyle W} \approx 4.3$ or by powers of $\MT$.  
The general expression for $\vec C_e^{(2)} (\mu)$ at the low scale is 
\beq \label{ce2}
\begin{split} 
\vec C_e^{(2)}(\mu) & = \eta \, \hat U^{(0)}_s (\mu, \muh) \vec
C_e^{(2)} (\muh) + \eta^2 \, \hat U_e^{(0)} (\mu, \muh) \vec C_s^{(2)}
(\muh) \\[1mm] 
& + \hat U^{(1)}_s (\mu, \muh) \vec C_e^{(1)} (\muh) + \eta \, \hat
U_e^{(1)} (\mu, \muh) \vec C_s^{(1)} (\muh) + \hat U_e^{(2)} (\mu,
\muh) \vec C^{(0)}_s (\muh) \, .  
\end{split}
\eeq
The second, third, and fourth term in this equation are known
completely, from \cite{Bobeth:1999mk} and the above discussion. The
first term is partially known: in particular $C_{e, 7}^{(2)} (\muh)$
and $C_{e, 8}^{(2)} (\muh)$ are known from \cite{ewnoi}, while only the
part of $C_{e,9}^{(2)} (\muh)$ and $C_{e,10}^{(2)} (\muh)$
proportional to $\MT^4$ is known from \cite{Buchalla:1997kz}. 
We also include in the following the leading contributions to
$C_{e,3}^{Q (2)} (\muh)$--$C_{e,6}^{Q (2)} (\muh)$ calculated in 
\cite{Bobeth:1999mk, Buras:1999st}\footnote{Since the calculation of
\Ref{Buras:1999st} has been performed in a different operator basis
one has to perform a non-trivial change of scheme. See \cite{future} for
details.}. The last term in \Eq{ce2} would require the knowledge of
the three-loop $\ord(\aem \as^2)$ ADM and will be neglected in what
follows. Again, rather than giving analytic expressions for these NNLO
corrections, we will only give some approximate formulas and refer to
a forthcoming publication \cite{future} for more details. We use the
results of \cite{Bobeth:1999mk, ewnoi, Buras:1999st, Buchalla:1997kz}
for the $\ord (\aem \as)$ initial conditions of the relevant Wilson
coefficients and find for the correction to $C_7 (\mu)$   
\beq \label{c7e2}
\begin{split}
C_{e,7}^{(2)} (\mu) & \approx 
2.9457 \, e^{-0.4323 \, \eta} \,\eta^{1.3783} x_t^{0.9498} + 1.2160 \,
e^{-5.4241 \, \eta} \, \eta^{2.1365} \, x_t^{-0.5479}  
\\[1mm]
& - 6.8887 \times 10^{10} \, e^{-34.605 \, \eta} \, \eta^{13.470} \, 
x_t^{0.9677} + 50.557 \, e^{-5.9375 \, \eta} \, \eta^{2.6489} \, 
x_t^{-0.0736} \\[1mm] 
& \approx  4.0718 + 0.0077 - 0.4315 + 0.3522 \, . 
\end{split}
\eeq 
The analogous expression for $C_9 (\mu)$ reads  
\beq \label{c9e2}
\begin{split}
C_{e,9}^{(2)} (\mu) & \approx 
2.3684 \times 10^{11} \, e^{-31.006 \, \eta} \, \eta^{12.546} \,
x_t^{-0.9681} - 209.52 \, e^{-4.6671 \eta}\, \eta^{2.3626}\,
x_t^{-1.7441} \\[1mm]     
& + 519.80 \, e^{-6.5763 \, \eta} \, \eta^{3.1984} \, x_t^{0.4666} + 
94.434 \, e^{-4.8623 \, \eta} \, \eta^{1.1723} \, x_t^{0.5775} \\[1mm]
& \approx 1.1803 - 0.3190 + 4.0021 + 7.2083 \, . 
\end{split}
\eeq 
Finally, for the $\ord (\aem \as)$ correction to $C_{10} (\mu)$ we get
\beq \label{c10e2}
\begin{split}
C_{e,10}^{(2)} (\mu) & \approx 0.3106 \, x_t^{2.6056}
- 242.56 \, e^{-4.7109 \, \eta} \, \eta^{1.5336} x_t^{-0.7682} \\[1mm]
& + 80.747 \, e^{-4.9425 \, \eta} \, \eta^{1.2373} \, x_t^{0.7343}
\\[1mm]  
& \approx 13.144 - 2.3657 + 7.1098 \, . 
\end{split}
\eeq
In the last equation the contribution of the third term in \Eq{ce2} is
exactly zero. In \Eqs{c7e2}, \eq{c9e2} and \eq{c10e2} we have set 
$\MH = 115 \GeV$ and the last line
always corresponds to the numerical value of the four individual
contributions to \Eq{ce2}, employing $\eta = 0.56$, $\MT(\MT) = 165
\GeV$, and setting $\muh = \MW$ and $\mut = \MT$. 
For what concerns $C_7 (\mb)$, these $\ord (\aem \as)$ corrections are 
twice as large as the $\ord (\aem)$ correction, and lead to a shift of
$0.00004$. On the other hand, in the case of $C_9 (\mb)$ and $C_{10} 
(\mb)$ they are much smaller than the corrections of $\ord (\aem)$, as
they amount to only $0.00016$ and $0.00018$. We stress that they are
{\it incomplete} and we study them  to check the potential size of the
higher order contributions. 
If we add these partial higher order
results to the complete leading and next-to-leading electroweak
effects considered above, the values of the Wilson coefficients of
$Q_7$, $Q_9$ and $Q_{10}$ are changed by $-1.1 \%$, $-2.9 \%$ and
$+2.3 \%$, respectively.   

At this point we also note that the  electroweak corrections
to $C_{10} (\mub)$ depend rather strongly on the $t$-quark mass
renormalization scale $\mu_t$. Using $\MT(\MW) = 175 \GeV$ instead of
$\MT(\MT) = 165 \GeV$, 
which usually leads to larger higher order QCD corrections,  one finds
a correction of $+5.3 \%$. 
This scheme dependence
stems from the parts of $C_{e,10}^{(2)} (\muh)$ proportional to
$\MT^4$ computed in \cite{Buchalla:1997kz}, which we have evaluated in
the $\MSbar$ scheme for the $t$-quark mass. On the other hand, the
$\mut$-dependence of the QED corrections to $C_7 (\mub)$ and $C_9  
(\mub)$ is very weak.    

 Let us now verify that the RGE has indeed preserved the correct
normalization of the electroweak coupling for the matching 
contributions to $Q_9$ and $Q_{10}$, as anticipated at the beginning
of this section. To this end, we first consider the effect of the 
$\beta_e^{(0)}$ terms in the ADM of \Eq{eq:gammae0}. Using the 
explicit form of $\hat U_e^{(0)} (\mu, \muh)$, one sees that they
modify the contribution of $C_9(\muh)$  to the low-energy Wilson 
coefficients as follows  
\beq 
\begin{split} \label{eq:beta0terms}
U_{s,99}^{(0)}(\mu,\muh) C_9 (\muh) & ~\to~ \left [ 1 - \aem
\f{\beta_e^{(0)}}{\beta_s^{(0)}} \left ( \f{1}{\as(\mu)} - 
\f{1}{\as(\muh)} \right ) \right ]
U_{s,99}^{(0)}(\mu,\muh) C_9(\muh) \\[2mm] 
& ~\approx~ \f{\aem(\muh)}{\aem(\mu)} U_{s,99}^{(0)} (\mu, \muh)
C_9(\muh) \, ,   
\end{split}
\eeq
where the equality holds up to small higher order terms in $\aem$
and up to subleading logarithms in $\as$.  Indeed, the consideration of
subleading QCD effects, like those contained in $\hat U_e^{(1)} (\mu,
\muh)$, shows that QCD does not affect the proper normalization of the
electroweak coupling, as anticipated above.
Equation \eq{eq:beta0terms} and the analogous relation for $C_{10} (\muh)$
demonstrate that the physically motivated normalization of the electroweak
coupling is preserved by the RG evolution even 
for our choice of normalization of the operators:
since the matrix element is
multiplied by $\aem(\mu)$, the rescaling factor in front of
$C_9(\muh)$ brings the normalization of the electromagnetic coupling
back to the electroweak scale. All other large logarithms 
described by the QED--QCD evolution matrices $\hat U_e^{(i)} (\mu,
\muh)$ \ are solely induced by the running of the operators.

While $\hat U^{(0)}_s (\mu, \muh)$ and $\hat U^{(0)}_e (\mu, \muh)$
are scheme-independent, $\hat U^{(1)}_s (\mu, \muh)$ and $\hat
U^{(1)}_e (\mu, \muh)$ depend on the renormalization scheme in a way
that is detailed for example in \Ref{Buras:1999st}. The scheme
dependence is canceled by the $\ord (\as)$ and $\ord (\aem)$ matrix
elements of the effective operators. In fact, any calculation that
goes beyond leading logarithms requires the knowledge of the one-loop matrix
elements. However, we have seen above that the dominant effects
originate at the weak scale and have often  to do with  $\ord
(\aem/\sin^2 \theta_{\scriptscriptstyle W})$ contributions to the
matching conditions. These contributions are all scheme-independent.
One could  also  absorb some contributions to the QED matrix elements
by  replacing  $\aem(\mu)$ with $\aem(\mll)$, as discussed at the
beginning of this section. However, numerically this would lead to a
small difference and we will avoid it. We therefore expect our
incomplete calculation to capture the dominant electroweak effects.  
Moreover, any complete calculation of the QED matrix elements
should take into account real photon radiation, which is very
sensitive to the details of the experimental set-up and requires a
dedicated study at the experimental level.   

\section{Phenomenology}
\label{sec:PH}

We are now ready to combine all our results and update the prediction
for $\BRll$. We first write the integrated branching ratio as in
\cite{Bobeth:1999mk, extraNNLO}:
\beq \label{br}
\BRll = \BR[\BtoXclnu] \int_{0.05}^{0.25} \! d {\hat s} \, 
\f{1}{\Gamma[\BtoXclnu]}\f{d \Gamma[\BtoXslpluslminus]}{d {\hat s}} \,
,  
\eeq
where
\beq \label{dgds}
\begin{split}
\f{d \Gamma[\BtoXslpluslminus]}{d {\hat s}} & = \f{\GF^2 m_{b, {\rm
pole}}^5 \left | V_{ts}^* V_{tb} \right |^2}{48 \pi^3}  \left (
\f{\aem (\mb)}{4 \pi}\right )^2 (1 - {\hat s})^2 \, \Bigg \{
\left (  4 + \f{8}{{\hat s}} \right ) \Big | \widetilde{C}_{7,\BR}^{\rm
eff}  (\hat s) \Big |^2 \\   
& \hspace{-3.0cm} + (1 + 2 {\hat s}) \left ( \Big |
\widetilde{C}_{9, \BR}^{\rm eff} ({\hat s}) \Big |^2+ \Big |
\widetilde{C}_{10, \BR}^{\rm eff}({\hat s}) \Big |^2 \right) + 12 {\, \rm 
Re} \left ( \widetilde{C}_{7, \BR}^{\rm eff} (\hat s)
\widetilde{C}_{9, \BR}^{\rm eff} (\hat s)^* \right )+ \f{d
\Gamma^{\rm Brems}}{d\hat s} \Bigg \} \, ,  
\end{split}
\eeq
and the effective Wilson coefficients include all real and virtual
QCD corrections \cite{extraNNLO} as indicated by the index $\BR$, whereas
the last term encodes finite bremsstrahlungs corrections computed in the
second paper of \Ref{extraNNLO}. At
$\ord (\aem)$ the definition of the effective Wilson coefficients is
affected by the presence of the   
electroweak penguin operators $Q_3^Q$--$Q_6^Q$ and of $Q_1^b$ --- for
a detailed discussion we refer to \cite{future}. 
In order to cancel the strong dependence on the bottom quark mass in
the factor $m_b^5$, it is customary to normalize $\BRll$ to the 
experimental value of the Branching Ratio (BR) for the inclusive 
semileptonic decay $\BR[\BtoXclnu]$. However, this normalization
introduces a strong dependence on the charm quark mass, which is not 
known very accurately. The ensuing uncertainty is about $8 \%$ 
\cite{ghinculov_last}. An alternative procedure \cite{gambino-misiak,
Chankowski:2003wz} consists in normalizing the $\BtoXslpluslminus$
decay width to $\Gamma[\BtoXulnu]$, and then to express $\BR[\BtoXulnu]$
in terms of $\BR[\BtoXclnu]$ and of the ratio   
\beq
C = \left | \f{V_{ub}}{V_{cb}} \right|^2
\f{\Gamma[\BtoXclnu]}{\Gamma[\BtoXulnu]} \, ,  
\eeq
which can be computed with better accuracy \cite{gambino-misiak}. In
other words, we will use 
\beq \label{brx}
\BRll = \f{\BR[\BtoXclnu]}{C} \left | \f{V_{ub}}{V_{cb}} \right|^2 \, 
        \int_{0.05}^{0.25} \! d \hat s \,
        \f{1}{\Gamma[\BtoXulnu]} \f{d \Gamma[\BtoXslpluslminus]}{d \hat s} \, .
\eeq
The main uncertainty in $C$ comes from non-perturbative effects. Using
an updated determination of the heavy quark effective theory parameter
$\lambda_1 = (-0.31 \pm 0.10) \GeV^2$ \cite{YR}, and following otherwise
\Ref{gambino-misiak}, we obtain $C = 0.581 \pm 0.017$. For the
inclusive semileptonic $b \to c$ branching ratio we use
$\BR_{sl} \equiv \BR[\BtoXclnu] = ( 10.74 \pm 0.24 ) \%$ obtained 
from the latest Heavy Flavor Averaging Group 
results \cite{HFAG} after subtracting the $b \to u$ transition. For
the Cabbibo-Kobayashi-Maskawa (CKM) factor 
$|V_{ts}^* V_{tb}|^2/|V_{cb}|^2$ we use $0.969 \pm
0.005$ \cite{YR}, but we actually perform the calculation keeping top, charm
and up quark sectors separated \cite{Bobeth:1999mk}. Finally, we
employ the $c$- and $b$-quark pole masses, $m_{c, \rm pole} = ( 1.4
\pm 0.2 )\GeV$ and $m_{b, \rm pole}=(4.80\pm 0.15)\GeV$.

\begin{figure}[t]
\hspace{-0.65cm}
\scalebox{0.4}{\includegraphics{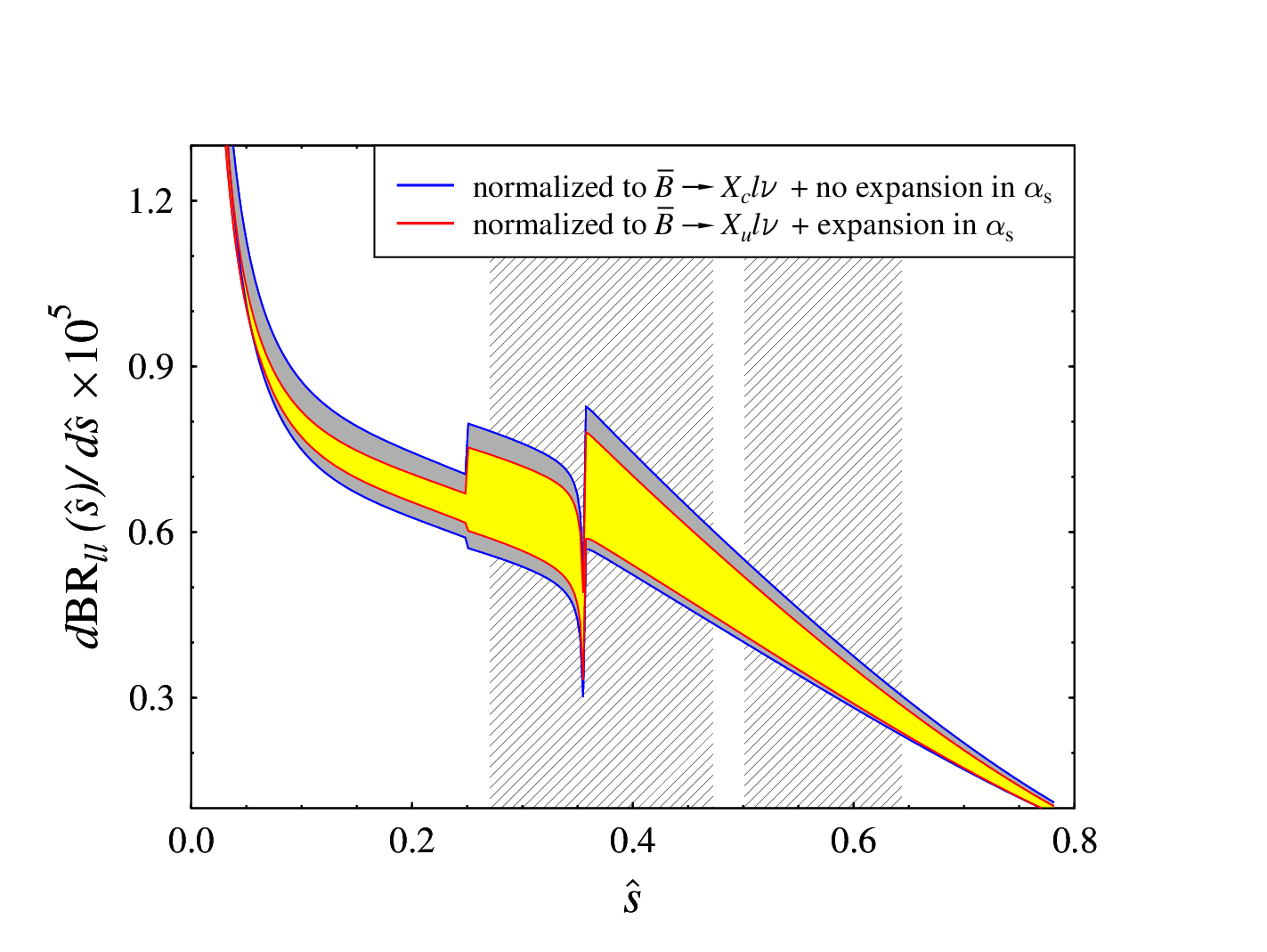}}
\hspace{-1.0cm}
\scalebox{0.4}{\includegraphics{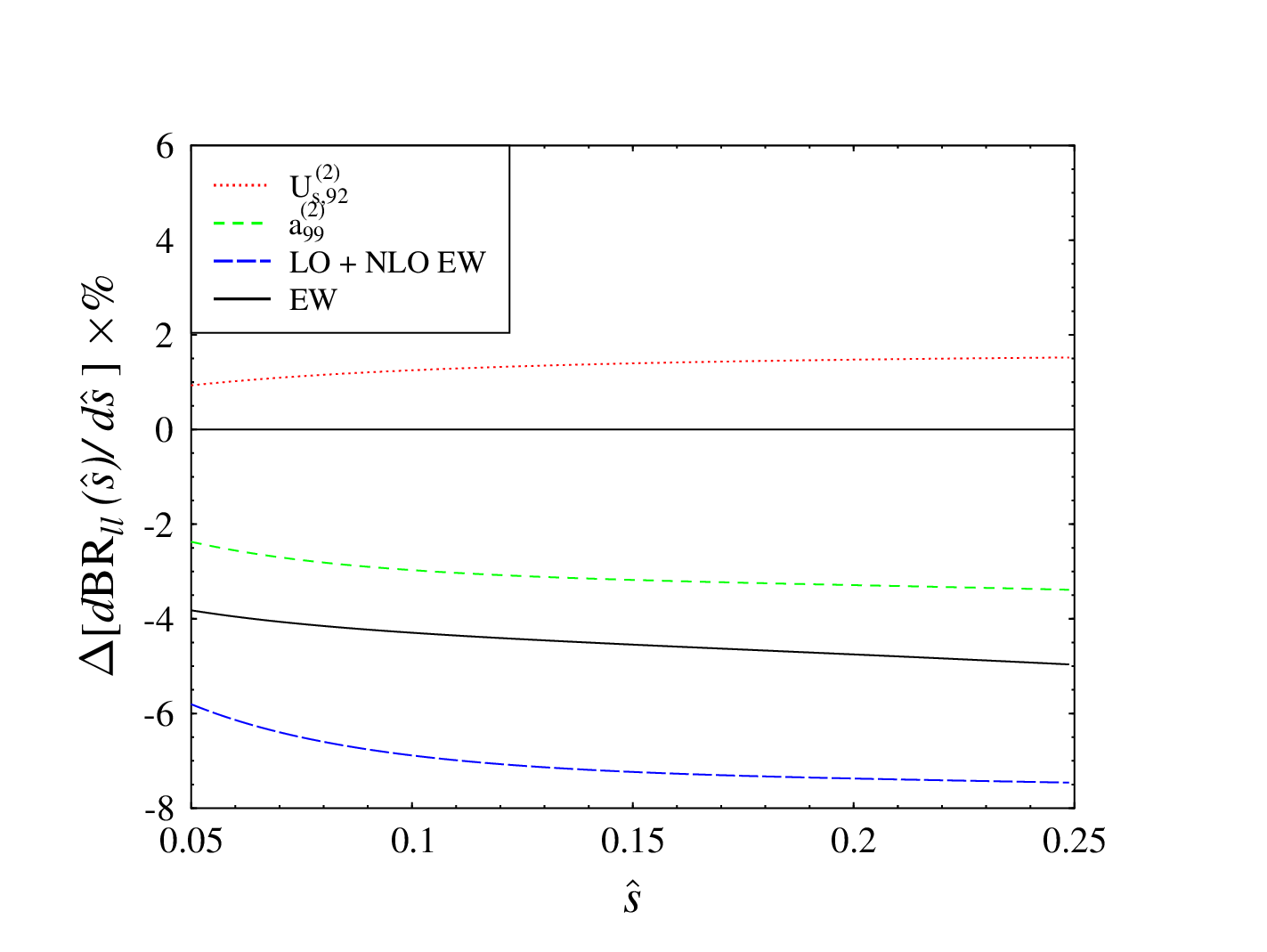}}
\vspace{-0.75cm}
\caption{\sf Scale dependence of the differential $\BRll$ using the
different normalizations of \Eqsand{br}{brx} in the partial NNLO QCD
approximation (lhs). The discontinuity at the end of the low-$\hat 
s$ region reflects the fact that the $\ord (\as^2)$ matrix elements of
$Q_1$ and $Q_2$ have not been implemented for $\hat s > 0.25$
\cite{ginonew}, while the divergence in the $J/\psi$ region is due to
the $1/\mc^2$ corrections. Furthermore the relative size of the NNLO
and electroweak (EW) effects introduced in this paper are shown when
{\it sequentially} included (rhs).}   
\label{fig:spectrum}
\end{figure}

The use of the $b$-quark pole mass in \Eq{dgds} and in the calculation
of  the semileptonic width leads to large perturbative QCD corrections
both in the numerator and denominator of \Eqsand{br}{brx}. 
Since this is an artifact of the choice of scheme for the bottom quark
mass, they clearly tend to cancel in the ratio. As a second improvement
with respect to previous analyses, we  keep terms through  $\ord (\as^2)$
in the denominator and expand the ratio in \Eq{brx} in
powers of $\as$. By making explicit the cancellation of large
contributions, we improve the convergence and stability of the
perturbative series, as we have explicitly checked. Like in previous
analyses, due to the peculiarity of the perturbative expansion for
$\btoslpluslminus$, we retain some large and scheme-independent
higher order term in the amplitude squared: for instance the terms 
$|\widetilde{C}^{\rm eff}_{7} (\hat s)|^2$ and $|\widetilde{C}^{\rm 
eff}_{10} (\hat s)|^2$ in \Eq{dgds} are expanded up to $\ord(\as)$. 
In addition all known $\ord(\as)$ terms of $|\widetilde{C}^{\rm
eff}_{9} (\hat s)|^2$ and ${\rm Re} ( \widetilde{C}_7^{\rm eff} (\hat
s) \widetilde{C}_9^{\rm eff} (\hat s)^* )$ are kept.
Our procedure differs slightly from the one proposed in the
first paper of \cite{extraNNLO}, which consists in absorbing the
numerically small coefficient $C_{s,9}^{(0)} (\mub)$ into
$C_{s,9}^{(1)} (\mub)$, and in subsequently performing the $\as$
expansion of the numerator
$d \Gamma[\BtoXslpluslminus]/d \hat s$ only. In the 
partial NNLO approximation, that is, 
setting $U_{s, 92}^{(2)}(\mub,\muh)$ and $a_{99}^{(2)} (\hat s)$ to zero, 
this alternative method yields
somewhat lower values for the normalized differential rate, well
within the scale dependence of our results.

We also include power corrections of $\ord(1/\mb^2)$ \cite{powercorrMB}
and $\ord(1/\mc^2)$ \cite{powercorrMC}  to the
differential decay rate of $\BtoXslpluslminus$ as well as
$\ord(1/\mb^2)$ corrections to $\Gamma[\BtoXclnu]$ and
$\Gamma[\BtoXulnu]$ --- see for example \cite{semiMB} --- and expand
the normalized differential decay rates given in \Eqsand{br}{brx} in
inverse powers of $\mb$ and $\mc$. 

\begin{table}[t] 
\begin{center}
\begin{tabular}{|l|c|c|c|c|c|}
\hline 
& & & & & \\[-4mm]
&partial NNLO & $U_{s, 92}^{(2)}$ & $a_{99}^{(2)}$ & $\text{LO} +
\text{NLO EW}$ 
& EW \\    
& & & & & \\[-4mm]
\hline 
& & & & & \\[-4mm]
$\BRll$ as in \eq{brshat}&
$1.509^{+0.035}_{-0.088}$ & $1.529^{+0.030}_{-0.081}$ & 
$1.463^{+0.038}_{-0.080}$ & $1.403^{+0.056}_{-0.087}$ & 
$1.442^{+0.054}_{-0.069}$ \\ 
& & & & & \\[-4mm]
\hline 
& & & & & \\[-4mm]
$\BRll$ as in \eq{brmll}&
$1.646^{+0.038}_{-0.095}$ & $1.668^{+0.032}_{-0.087}$ & 
$1.597^{+0.041}_{-0.087}$ & $1.532^{+0.060}_{-0.094}$ & 
$1.574^{+0.059}_{-0.075}$ \\ [0.5mm]
\hline
\end{tabular}
\end{center}
\caption{\sf $\BRll$ integrated over the low-$\hat s$ and low-$q^2$
region in units $10^{-6}$. The different corrections are added in sequence.} 
\label{tab:brs} 
\end{table}

The  partial NNLO results before inclusion of the two
contributions discussed  
in Section 2 are shown in the left plot of
\Fig{fig:spectrum}, where the bands include only the residual scale
dependence, calculated performing a scan in the ranges: $[40 \,
\text{--} \, 120] \GeV$ for the matching scale in the charm and up
sector, $[60 \, \text{--} \, 180] \GeV$ for the matching scale in the 
top sector, $[2.5 \, \text{--} \, 10] \GeV$ for the low-energy
renormalization scale. In the last case, the central value is set to 
$5 \GeV$. As can be nicely
seen by comparing the gray with the yellow band, the residual scale
dependence for low $\hat s$ is about $50 \%$ smaller in our approach than 
without the two improvements introduced between \Eqsand{dgds}{brx}. On the other
hand, the scale uncertainty is practically unaffected by the new
NNLO contributions --- not shown in \Fig{fig:spectrum}.  


The results for the integrated BR are given in the first line of
\Tab{tab:brs}, where the first column refers to the partial NNLO
calculation according to \cite{extraNNLO} but implemented as described
above, the second column to the same calculation supplemented by our
result for $U_{s,92}^{(2)}(\mub, \muh)$, the third column to the
complete NNLO, that is, including the two-loop QCD matrix element
$a_{99}^{(2)} (\hat s)$. The fourth column refers to the NNLO
calculation including LO and NLO electroweak effects, and the fifth
column to the QCD computation supplemented by the electroweak effects
through NNLO calculated in the previous section. Furthermore, in the
last two columns the dominant QED corrections to the semileptonic
decay amplitude \cite{Sirlin:1981ie} have been included.   
If we had only  rescaled the matching conditions for 
$Q_9$ and $Q_{10}$ and for the penguin operators by a factor
$\aem(\MW)/\aem(\mb)$ as discussed in the previous
section, the shift would be close to $+7 \%$.  
This effect is overcompensated by other electroweak corrections of
around $- 7\%$ and by the QED correction to the semileptonic rate,
which by itself amounts to around  $-1.5 \%$. We would like to
stress that this is an unexpected accidental cancellation. The right 
plot of \Fig{fig:spectrum} shows $\Delta[d\BRll(\hat{s})/d\hat s]$ ---
the differential BR normalized to the partial NNLO result --- where we
have sequentially included the corrections listed above.  

Including the main  parametric uncertainties, our final result reads  
\begin{align} \label{brshat}
& \BRll \left ( 0.05 \le \hat s \le 0.25 \right ) = \\[1mm] 
& \hspace{1cm} \left [ 
  1.442
  \pm^{0.098}_{0.092} |_{\MT}
  \pm^{0.054}_{0.069} |_{\rm scale} 
  \pm 0.041_{C}
  \pm 0.032_{\BR_{sl}}
  \pm^{0.001}_{0.009}|_{\mc} 
  \pm 0.002_{\mb}
  \right ] \times 10^{-6}
\, . \non 
\end{align}
Here the whole dependence on $\lambda_1$ is absorbed into the factor
$C$. As already in the case of $C$, we neglect effects suppressed by
three powers of the heavy quark masses in $\BtoXslpluslminus$
\cite{powercorrMB3}, which in the low-$\hat s$ region should not
exceed  $2\%$ anyway. The uncertainty related to the CKM matrix
elements is below $1 \%$. To be conservative, one could also consider 
an additional $2\%$ uncertainty due to unknown electroweak effects.
Combining in quadrature all uncertainties given in \Eq{brshat} with 
the latter three, the total error is about $9\%$, very
similar to the one for inclusive radiative decays
\cite{gambino-misiak}.

Though useful for comparison with previous analyses
\cite{extraNNLO}, the BR integrated in the region 
$0.05 \le \hat s \le 0.25$ is an idealization. A quantity  closer to
experiment is the BR integrated in the region $1 \GeV^2 \le q^2 \le
6 \GeV^2$ \cite{ghinculov_last}. The results for the integrated BR
employing this range are given in the second line of  
\Tab{tab:brs} together with their scale dependence. 
Unfortunately, the choice of the integration variable 
introduces an enhanced dependence on the bottom quark mass. In 
our calculation we have employed the $b$-quark pole mass, which is
subject to a much larger uncertainty than an appropriate 
short-distance $b$-quark mass. For example 
the $1 S$ or the kinetic low-energy $b$-quark mass \cite{YR} are known
to better than $50 \MeV$. Our final result is   
\begin{align} \label{brmll}
& \BRll \left (1 \GeV^2 \le q^2 \le 6 \GeV^2 \right ) = \\[1mm]
& \hspace{1cm} \left [
  1.574
  \pm^{0.106}_{0.100} |_{\MT}                       
  \pm^{0.072}_{0.067} |_{\mb} 
  \pm^{0.059}_{0.075} |_{\rm scale} 
  \pm 0.045_{C}
  \pm 0.035_{\BR_{sl}}
  \pm^{0.001}_{0.013} |_{\mc} 
  \right ] \times 10^{-6} \, . \non 
\end{align}
The total error is about $10 \%$ and is again dominated by
the uncertainty on the top quark mass, which will soon be reduced by a
factor of two by CDF and D0 at the Tevatron. 
Moreover, the substantial error from the $b$-quark mass
is an artifact of the employed scheme, which can be reduced by a
factor of three or even more by changing the renormalization scheme of
the $b$-quark mass. Hence it should not be interpreted as a
limitation. 

For what concerns the FB asymmetry, we have
investigated the impact of the choice of normalization 
and of the electroweak corrections on the  so-called unnormalized asymmetry
defined as in \cite{extraNNLO}
\beq \label{afbxc}
A_{\rm FB}(\hat s)=\f{\BR[\BtoXclnu]}{\Gamma[\BtoXclnu]}
\int_{-1}^1 \! d\cos\theta_\ell \, 
\f{d^2 \Gamma[\BtoXslpluslminus]}{d\hat s \,d\cos\theta_\ell}
\,{\rm sgn}(\cos\theta_\ell) \, ,  
\eeq
where 
\beq \label{afb}
\begin{split}
\int_{-1}^1 \! d\cos\theta_\ell \, \f{d^2
\Gamma[\BtoXslpluslminus]}{d\hat s \,d\cos\theta_\ell} \,{\rm
sgn}(\cos\theta_\ell) & = 
\f{\GF^2 m_{b, {\rm pole}}^5 \left | V_{ts}^* V_{tb} \right |^2}{48
\pi^3} \left ( \f{\aem (\mb)}{4 \pi}\right )^2 (1 - {\hat s})^2 \\
& \hspace{-6.5cm} \times \Bigg \{ \! -6 \, {\rm Re} \left (
\widetilde{C}^{\rm eff}_{7, {\rm FB}} (\hat s) \widetilde{C}^{{\rm
eff}}_{10, {\rm FB}} (\hat s)^\ast \right ) - 3 \hspace{0.1mm} \hat{s}
\, {\rm Re} \left ( \widetilde{C}^{\rm eff}_{9, {\rm FB}} (\hat s)
\widetilde{C}^{{\rm eff}}_{10, {\rm FB}} (\hat s)^\ast \right ) +
A_{\rm FB}^{\rm Brems} (\hat s) \Bigg \} \, , 
\end{split}
\eeq
and the effective Wilson coefficients include all real and virtual
corrections calculated in \cite{extraNNLO} as indicated by the
subscript $\rm FB$. The last term of \Eq{afb} encodes finite
bremsstrahlungs corrections evaluated in the last reference of
\cite{extraNNLO} which we have not included\footnote{These corrections
turn out to be below $1 \%$.}. At $\ord (\aem)$ the definition of the
effective Wilson coefficients is affected by the presence of the
electroweak penguin operators $Q_3^Q$--$Q_6^Q$ and of $Q_1^b$ --- for
a detailed discussion we again refer to \cite{future}. 
\begin{table}[t] 
\begin{center}
\begin{tabular}{|l|c|c|}
\hline 
& & \\[-4mm]
& $\hat s_0$ without $\ord (1/m_{b,c}^2)$ & $\hat s_0$ with $\ord
(1/m_{b,c}^2)$ \\ 
& & \\[-4mm]
\hline 
& & \\[-4mm]
unexpanded $A_{\rm FB} (\hat s)$ as in \eq{afbxc} & {
$0.160^{+0.008}_{-0.005}$} & { $0.167^{+0.008}_{-0.006}$} \\  
& & \\[-4mm]
\hline 
& & \\[-4mm]
expanded $A_{\rm FB} (\hat s)$ as in \eq{afbxu} &
{ $0.151^{+0.005}_{-0.004}$} & { $0.156^{+0.005}_{-0.005}$} \\
& & \\[-4mm]
\hline 
& & \\[-4mm]
expanded ${\bar A}_{\rm FB} (\hat s)$ as in \eq{afbbar} & 
{ $0.157^{+0.005}_{-0.009}$} & { $0.163^{+0.007}_{-0.010}$} \\
& & \\[-4mm]
\hline 
& & \\[-4mm]
unexpanded $A_{\rm FB} (\hat s)$ as in \eq{afbxc} + { EW}& { 
$0.164^{+0.004}_{-0.004}$} & { $0.170^{+0.004}_{-0.005}$} \\ 
& & \\[-4mm]
\hline 
& & \\[-4mm]
expanded $A_{\rm FB} (\hat s)$ as in \eq{afbxu} + { EW}&
{ $0.154^{+0.004}_{-0.004}$} & { $0.159^{+0.004}_{-0.004}$} \\
& & \\[-4mm]
\hline 
& & \\[-4mm]
expanded ${\bar A}_{\rm FB} (\hat s)$ as in \eq{afbbar} + { EW} &
{ $0.160^{+0.006}_{-0.009}$} & { $0.166^{+0.007}_{-0.010}$} \\[0.5mm]
\hline 
\end{tabular}
\end{center}
\caption{\sf Zero of FB asymmetry 
before and
after the inclusion of $1/\mb^2$ and $1/\mc^2$ power corrections.}   
\label{tab:afb} 
\end{table} 

Implementing all the relevant NNLO QCD corrections we obtain for the
position of the zero of the unnormalized FB asymmetry \eq{afbxc}
the values given in the first
column of the first line of \Tab{tab:afb},  in
agreement with \cite{extraNNLO, ginonew}. While in
Refs.~\cite{extraNNLO, ginonew} only the low-energy  renormalization
scale was varied, the errors in \Tab{tab:afb} also include the
matching scale dependence. 

Since the position of the zero of the FB asymmetry is known to be
especially sensitive to physics beyond the SM, a comprehensive study
of the residual theoretical error attached to it is important. In
order to get a grasp for the size of higher order corrections, we
consider the normalization to the charmless semileptonic decay rate   
\beq \label{afbxu}
A_{\rm FB}(\hat s)=\f{\BR[\BtoXclnu]}{\Gamma[\BtoXulnu]} \f{1}{C} 
\left | \f{V_{ub}}{V_{cb}} \right |^2 
\int_{-1}^1 \! d\cos\theta_\ell \, 
\f{d^2 \Gamma[\BtoXslpluslminus]}{d\hat s \,d\cos\theta_\ell}
\,{\rm sgn}(\cos\theta_\ell) \, ,  
\eeq
in combination with $C$ and perform an expansion of the ratio of
\Eq{afbxu} in $\as$ to improve the convergence of the perturbative
series. The corresponding result is displayed in the first entry of
the second line of \Tab{tab:afb}. Note that in this case the numerical
value obtained for $\hat{s}_0$ does {\it no longer} correspond to the
zero of the curly bracket in \Eq{afb}. The difference between the
values of $\hat s_0$ in the first and second line of \Tab{tab:afb} is
obviously due to different higher order terms and can give an
indication of the residual theoretical error.  

\begin{figure}[t]
\center{\scalebox{0.5}{\includegraphics{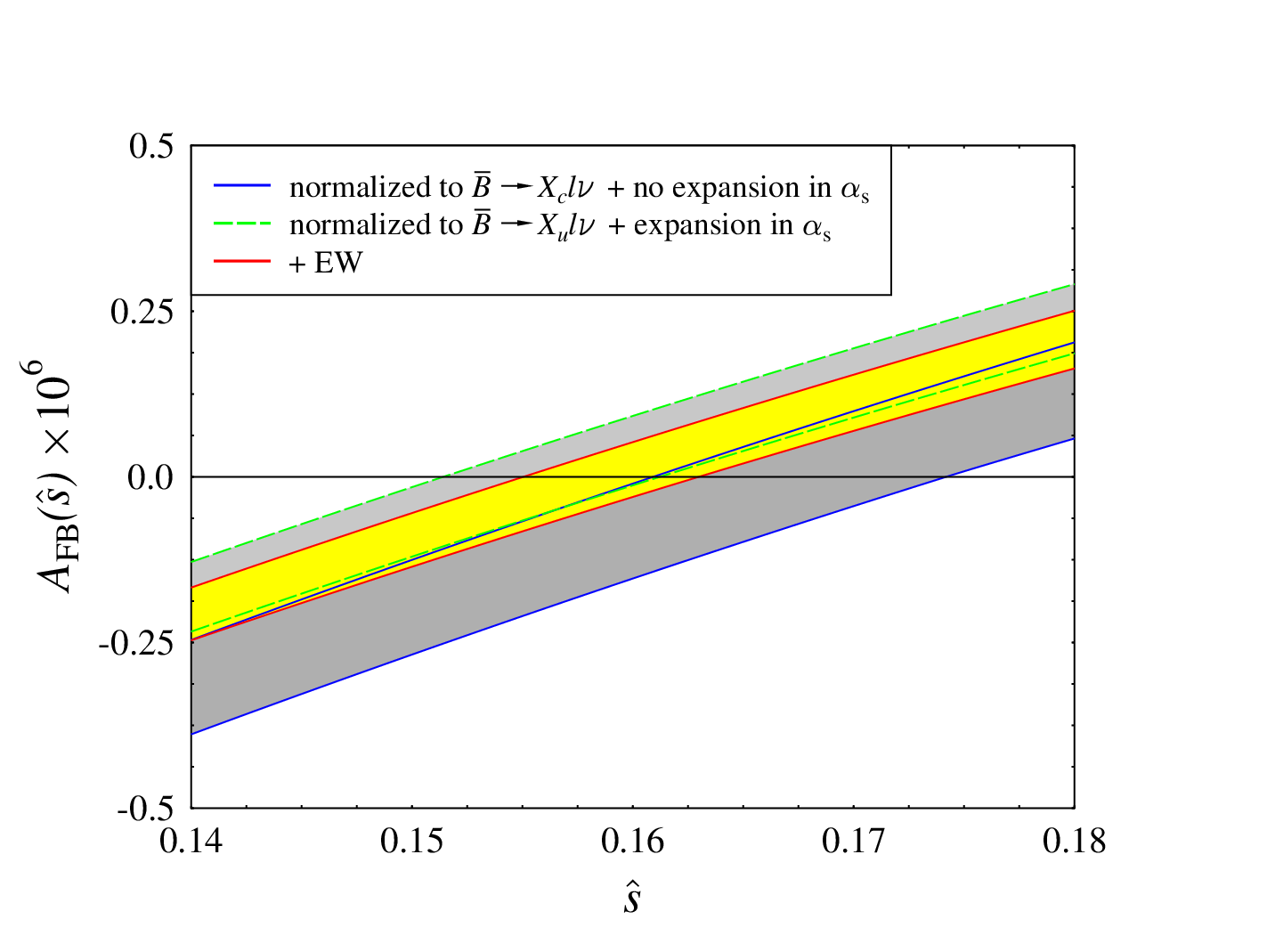}}}

\vspace{-2mm}
\caption{\sf 
Unnormalized FB asymmetry $A_{\rm FB} (\hat s)$ in the low-$\hat s$
region including $1/\mb^2$ and $1/\mc^2$ power corrections. Compared
is the NNLO QCD result for the normalization given in \Eq{afbxc} and 
the one given in \Eq{afbxu} performing a subsequent expansion in
$\as$. Furthermore we also display the result of \Eq{afbxu} after the
inclusion of electroweak effects (central yellow band).}  
\label{fig:as}
\end{figure}

We also consider the so-called normalized FB asymmetry, which is
certainly a quantity closer to experiment:       
\beq \label{afbbar}
\bar{A}_{\rm FB}(\hat s)=\f{1}{d \Gamma[\BtoXslpluslminus]/d \hat{s}}
\int_{-1}^1 \! d\cos\theta_\ell \, 
\f{d^2 \Gamma[\BtoXslpluslminus]}{d\hat s \,d\cos\theta_\ell}
\,{\rm sgn}(\cos\theta_\ell) \, .  
\eeq
Given the small value of the LO term of $d \Gamma[\BtoXslpluslminus]/d
\hat{s}$, a naive $\as$ expansion of this ratio is
problematic. However, the $\as$ expansion converges reasonably well if
the perturbative expansion is reorganized following the first paper of
\cite{extraNNLO}, namely absorbing $C_{s,9}^{(0)} (\mub)$ into
$C_{s,9}^{(1)} (\mub)$. With respect to the unexpanded form, which
obviously reproduces the value given in the first column of the first
line of \Tab{tab:afb}, the zero is shifted toward lower values, as
can be seen by comparing the first and the third entry in the first
column of that table. Finally, the next three entries of the first
column of \Tab{tab:afb} correspond to the results for $\hat s_0$ in
the three approaches described above after the inclusion of the
electroweak effects.

\begin{figure}[t]
\center{\scalebox{0.5}{\includegraphics{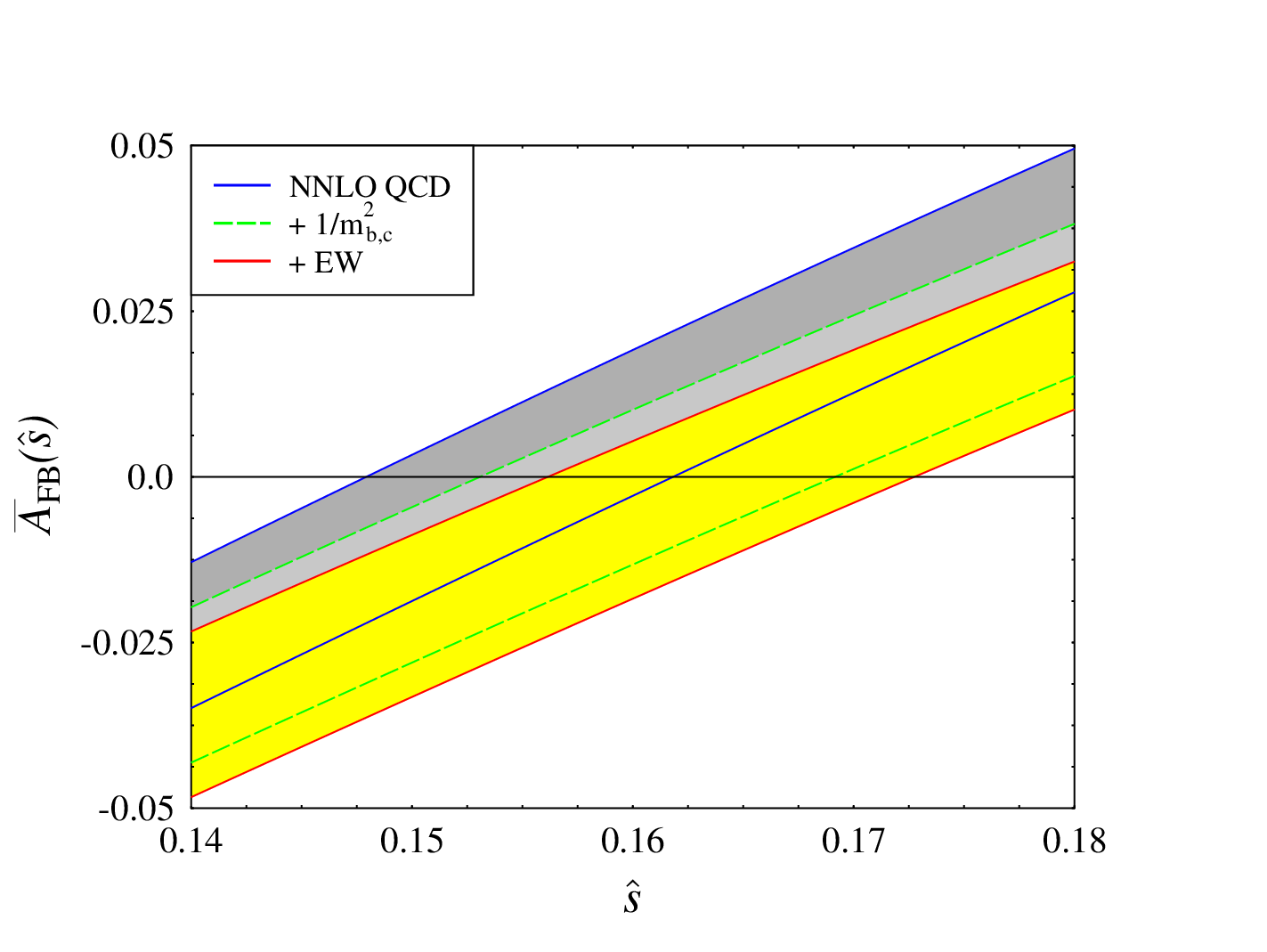}}}

\vspace{-2mm}
\caption{\sf  
Normalized FB asymmetry $\bar{A}_{\rm FB} (\hat s)$ in the low-$\hat
s$ region in the NNLO QCD approximation as given in \Eq{afbbar}
without expansion in $\as$. Furthermore the effects of including power
corrections (central band) and higher order electroweak effects (lower
yellow band) are shown.} 

\label{fig:asbar}
\end{figure}
Unlike previous analyses, we also include power corrections of $\ord
(1/\mb^2)$ \cite{powercorrMB} and $\ord (1/\mc^2)$ \cite{powercorrMC}
in the computation of the zero of the FB asymmetry. In doing so we
always expand the asymmetry in powers of the heavy quark masses. The 
corresponding numerical values for $\hat s_0$ can be found in the
second column of \Tab{tab:afb}. Comparing the results in that table,
one observes that higher order electroweak effects amount to around
$+2 \%$, independent on which normalization is used and on whether an
expansion in $\as$ is performed or not. The $\ord (1/\mb^2)$ and $\ord
(1/\mc^2)$ corrections turn out to be at the level of $+4 \%$. The
unnormalized FB asymmetry $A_{\rm FB} (\hat s)$ calculated in the two
above approaches in the vicinity of the zero is shown in \Fig{fig:as},
whereas in \Fig{fig:asbar} we plot the unexpanded $\bar{A}_{\rm
FB}(\hat s)$, showing the importance of the power and electroweak
corrections.   

A glance at \Tab{tab:afb} and \Figsand{fig:as}{fig:asbar} shows that
the dependence of the zero on the specific method used to calculate it
is generally
larger than the scale dependence associated with it. This suggests
that the observed scale dependence alone
does {\it not} provide a reasonable estimate of higher order QCD and 
electroweak corrections, as pointed out already 
in the third paper of \cite{extraNNLO}. 
Following a reasoning similar as for $\BRll$, we adopt as the central
value for $\hat s_0$ the average of the two expanded results in
\Tab{tab:afb} and  assign to it a rather conservative total error of  
$6 \%$, that covers  the whole range of possible values:      
\beq \label{sbu}
\hat s_0 = 0.163 \pm 0.010_{\rm theory} \, .
\eeq
The uncertainties due to $\MT$ and the non-perturbative parameters
$\lambda_1$ and $\lambda_2$ are at the 0.1\% level or less. Converting
this result into a value of $q^2$ introduces again a large
uncertainty  due to the  bottom quark mass. We find    
\beq \label{qbu}
q^2_0 = \left ( 3.76  \pm {0.22}_{\rm theory}
\pm 0.24_{\mb} 
\right ) \GeV^2 \, ,     
\eeq
with a  total error close to $9 \%$. However, also in this case the error
due to the $b$-quark mass can be drastically reduced by performing the
calculation in a different mass scheme. 

\section{Summary}
Rare semileptonic $B$ decays are going to play an important role in
the search for new physics at the $B$ factories and upcoming flavor
physics experiments at the Tevatron and the LHC. From the theoretical
point of view the inclusive mode at low $q^2$ is particular
interesting, since it can be accurately computed in the SM. In this
paper we have first completed the NNLO QCD analysis of
$\BtoXslpluslminus$ with the calculation of the last two missing 
components, and then discussed electroweak effects. The new NNLO QCD
contributions are relatively small --- they amount to around $+1
\%$ and $-4 \%$ --- and tend to partially cancel each other. 
Our treatment of electroweak effects is complete only for what
concerns the leading logarithms of photonic origin but, pending an
exhaustive investigation of the missing matrix elements
\cite{future,misiaknew}, we expect it to capture all the dominant
contributions. The impact of electroweak effects depends on the scale
employed for $\aem (\mu)$ in the prefactor for $\BRll$. As a
result of accidental cancellations, they change the value of the
low-energy BR by only $-1.5 \%$ for $\mu = \ord (\mb)$, while if one
adopts $\mu = \ord (\MW)$ the effect is much more relevant, about
$-8.5 \%$. Our calculation drastically reduces the large uncertainty
of almost $8 \%$ related to higher order electroweak effects, which
has affected previous analyses. We have also calculated the FB
asymmetry. The location of its zero is shifted by approximately $+2
\%$ due to the electroweak corrections.  

We have also introduced a few improvements in the calculation of the  
decay rate in the region of low dilepton invariant masses, aimed at
reducing the theoretical error. 
After a careful study of the various sources of uncertainty, we
estimate a $9 \%$ to $10 \%$ error on our result for the BR in
the $\hat s = 0.05 \, \text{--} \, 0.25$ or $q^2 = 1 \text{--} \, 6
\GeV^2$ window, the dominant source of which is the uncertainty on the
top quark mass. For what concerns the zero of the FB asymmetry, our
estimate for the residual uncertainty is $6 \%$ to  $9\%$.   

\vspace{5mm}

\subsubsection*{Acknowledgments}

During the final stage of our work we have become aware of an ongoing
analysis \cite{misiaknew} of electroweak effects in
$\BtoXslpluslminus$ that overlaps in part with Section 3 of this 
paper. We are grateful to Mikolaj Misiak for very informative
discussions, for reading carefully a preliminary version of our 
manuscript, and for reminding us of the role of $Q_1^b$,
as well as to Tobias Huber and Enrico Lunghi for pointing out the 
normalization error in Eq. (5).
We would also like to thank Andrzej Buras and
Gino Isidori for many useful discussions. The work of C.~B.\ was supported
by the {\it Deutsche Forschungsgemeinschaft} (DFG) under contract
Bu.706/1-2 and by the G.I.F Grant No.~G-698-22.7/2001, and in part by
the U.S. Department of Energy under contract
No.~DOE-FG03-97ER40546. The work of M.~G.\ was supported by the {\it 
Deutsches Bundesministerium f\"ur Bildung und Forschung} under the
contract No.~05HT1WOA3. The work of P.~G.\ is supported by a Marie
Curie Fellowship, contract No.~HPMF-CT-2000-01048. Finally, the work
of U.~H.\ is supported by the U.S. Department of Energy under contract
No.~DE-AC02-76CH03000.

\section*{Appendix}

We report below the complete $\ord (\as)$,  $\ord (\as^2)$,  $\ord
(\aem)$ and $\ord (\aem \as)$ ADMs making explicit their dependence
on the coefficients of the QED and QCD $\beta$ function:  
\beq 
\beta_s = -\f{\gs^3}{16 \pi^2} \Bigg ( \beta_s^{(0)} + \f{\gs^2}{16
\pi^2} \beta_s^{(1)} + \f{e^2}{16 \pi^2} \beta_{se}^{(1)} \Bigg ) \, ,
\hspace{10mm} \beta_e = \f{e^3}{16 \pi^2} \Bigg ( \beta_e^{(0)} +
\f{\gs^2}{16 \pi^2} \beta_{es}^{(1)} \Bigg ) \, .     
\eeq
For five active quark and three active lepton flavors the values of
the $\MSbar$ $\beta$ coefficients are   
\beq
\beta_s^{(0)} = \f{23}{3} \, , \hspace{5mm} \beta_s^{(1)} = \f{116}{3}
\, , \hspace{5mm} \beta_{se}^{(1)} = -\f{22}{9}\, , \hspace{5mm}
\beta_e^{(0)} = \f{80}{9} \, , \hspace{5mm} \beta_{es}^{(1)} =
\f{176}{9} \, .  
\eeq


Adopting the ordering of operators as introduced in \Eq{list} the
matrices $\hat{\gamma}_s^{(0)}$ and $\hat{\gamma}_s^{(1)}$ take the
following form 
{%
\renewcommand{\arraycolsep}{2.5pt}
\beq \label{eq:gammas0}
\hat{\gamma}_s^{(0)} = 
\left (
\begin{array}{ccccccccccccccc}
-4 & {\scriptstyle \f{8}{3}} & 0 & {\scriptstyle -\f{2}{9}} & 0 & 0 &
0 & 0 & 0 & 0 & 0 & 0 & 0 & {\scriptstyle -\f{32}{27}} & 0 \\  
12 & 0 & 0 & {\scriptstyle \f{4}{3}} & 0 & 0 & 0 & 0 & 0 & 0 & 0 & 0 &
0 & {\scriptstyle -\f{8}{9}} & 0 \\  
0 & 0 & 0 & {\scriptstyle -\f{52}{3}} & 0 & 2 & 0 & 0 & 0 & 0 & 0 & 0
& 0 & {\scriptstyle -\f{16}{9}} & 0 \\  
0 & 0 & {\scriptstyle -\f{40}{9}} & {\scriptstyle -\f{100}{9}} &
{\scriptstyle \f{4}{9}} & {\scriptstyle \f{5}{6}} & 0 & 0 & 0 & 0 & 0
& 0 & 0 & {\scriptstyle \f{32}{27}} & 0 \\  
0 & 0 & 0 & {\scriptstyle -\f{256}{3}} & 0 & 20 & 0 & 0 & 0 & 0 & 0 &
0 & 0 & {\scriptstyle -\f{112}{9}} & 0 \\  
0 & 0 & {\scriptstyle -\f{256}{9}} & {\scriptstyle \f{56}{9}} &
{\scriptstyle \f{40}{9}} & {\scriptstyle -\f{2}{3}} & 0 & 0 & 0 & 0 &
0 & 0 & 0 & {\scriptstyle \f{512}{27}} & 0 \\  
0 & 0 & 0 & {\scriptstyle -\f{8}{9}} & 0 & 0 & 0 & -20 & 0 & 2 & 0 & 0
& 0 & {\scriptstyle -\f{272}{27}} & 0 \\  
0 & 0 & 0 & {\scriptstyle \f{16}{27}} & 0 & 0 & {\scriptstyle
-\f{40}{9}} & {\scriptstyle -\f{52}{3}} & {\scriptstyle \f{4}{9}} &
{\scriptstyle \f{5}{6}} & 0 & 0 & 0 & {\scriptstyle -\f{32}{81}} & 0
\\  
0 & 0 & 0 & {\scriptstyle -\f{128}{9}} & 0 & 0 & 0 & -128 & 0 & 20 & 0
& 0 & 0 & {\scriptstyle -\f{2768}{27}} & 0 \\  
0 & 0 & 0 & {\scriptstyle \f{184}{27}} & 0 & 0 & {\scriptstyle
-\f{256}{9}} & {\scriptstyle -\f{160}{3}} & {\scriptstyle \f{40}{9}} &
{\scriptstyle -\f{2}{3}} & 0 & 0 & 0 & {\scriptstyle -\f{512}{81}} & 0
\\  
0 & 0 & 0 & {\scriptstyle \f{4}{3}} & 0 & 0 & 0 & 0 & 0 & 0 & 4 & 0 &
0 & {\scriptstyle \f{16}{9}} & 0 \\  
0 & 0 & 0 & 0 & 0 & 0 & 0 & 0 & 0 & 0 & 0 & {\scriptstyle \f{32}{3} -
2 \beta_s^{(0)}} & 0 & 0 & 0 \\  
0 & 0 & 0 & 0 & 0 & 0 & 0 & 0 & 0 & 0 & 0 & {\scriptstyle -\f{32}{9}}
& {\scriptstyle \f{28}{3} - 2 \beta_s^{(0)}} & 0 & 0 \\  
0 & 0 & 0 & 0 & 0 & 0 & 0 & 0 & 0 & 0 & 0 & 0 & 0 & {\scriptstyle
-2 \beta_s^{(0)}} & 0 \\  
0 & 0 & 0 & 0 & 0 & 0 & 0 & 0 & 0 & 0 & 0 & 0 & 0 & 0 & {\scriptstyle
-2 \beta_s^{(0)}} 
\end{array}
\right ) \, , 
\eeq
}%
{%
\renewcommand{\arraycolsep}{0.0pt}
\beq \label{eq:gammas1}
\hat{\gamma}_s^{(1)} = 
\left (
\begin{array}{ccccccccccccccc}
{\scriptstyle -\f{355}{9}} & {\scriptstyle -\f{502}{27}} &
{\scriptstyle -\f{1412}{243}} & {\scriptstyle -\f{1369}{243}} &
{\scriptstyle \f{134}{243}} & {\scriptstyle -\f{35}{162}} & 0 & 0 & 0
& 0 & 0 & {\scriptstyle -\f{232}{243}} & {\scriptstyle \f{167}{162}} &
{\scriptstyle -\f{2272}{729}} & 0 \\  
{\scriptstyle -\f{35}{3}} & {\scriptstyle -\f{28}{3}} & {\scriptstyle
-\f{416}{81}} & {\scriptstyle \f{1280}{81}} & {\scriptstyle
\f{56}{81}} & {\scriptstyle \f{35}{27}} & 0 & 0 & 0 & 0 & 0 &
{\scriptstyle \f{464}{81}} & {\scriptstyle \f{76}{27}} & {\scriptstyle
\f{1952}{243}} & 0 \\  
0 & 0 & {\scriptstyle -\f{4468}{81}} & {\scriptstyle -\f{31469}{81}} &
{\scriptstyle \f{400}{81}} & {\scriptstyle \f{3373}{108}} & 0 & 0 & 0
& 0 & 0 & {\scriptstyle \f{64}{81}} & {\scriptstyle \f{368}{27}} &
{\scriptstyle -\f{6752}{243}} & 0 \\  
0 & 0 & {\scriptstyle -\f{8158}{243}} & {\scriptstyle -\f{59399}{243}}
& {\scriptstyle \f{269}{486}} & {\scriptstyle \f{12899}{648}} & 0 & 0
& 0 & 0 & 0 & {\scriptstyle -\f{200}{243}} & {\scriptstyle
-\f{1409}{162}} & {\scriptstyle -\f{2192}{729}} & 0 \\  
0 & 0 & {\scriptstyle -\f{251680}{81}} & {\scriptstyle
-\f{128648}{81}} & {\scriptstyle \f{23836}{81}} & {\scriptstyle
\f{6106}{27}} & 0 & 0 & 0 & 0 & 0 & {\scriptstyle -\f{6464}{81}} &
{\scriptstyle \f{13052}{27}} & {\scriptstyle -\f{84032}{243}} & 0 \\  
0 & 0 & {\scriptstyle \f{58640}{243}} & {\scriptstyle -\f{26348}{243}}
& {\scriptstyle -\f{14324}{243}} & {\scriptstyle -\f{2551}{162}} & 0 &
0 & 0 & 0 & 0 & {\scriptstyle -\f{11408}{243}} & {\scriptstyle
-\f{2740}{81}} & {\scriptstyle -\f{37856}{729}} & 0 \\  
0 & 0 & {\scriptstyle \f{832}{243}} & {\scriptstyle -\f{4000}{243}} &
{\scriptstyle -\f{112}{243}} & {\scriptstyle -\f{70}{81}} &
{\scriptstyle -\f{404}{9}} & {\scriptstyle -\f{3077}{9}} &
{\scriptstyle \f{32}{9}} & {\scriptstyle \f{1031}{36}} & 0 &
{\scriptstyle -\f{64}{243}} & {\scriptstyle -\f{368}{81}} &
{\scriptstyle -\f{24352}{729}} & 0 \\  
0 & 0 & {\scriptstyle \f{3376}{729}} & {\scriptstyle \f{6344}{729}} &
{\scriptstyle -\f{280}{729}} & {\scriptstyle \f{55}{486}} &
{\scriptstyle -\f{2698}{81}} & {\scriptstyle -\f{8035}{27}} &
{\scriptstyle -\f{49}{162}} & {\scriptstyle \f{4493}{216}} & 0 &
{\scriptstyle \f{776}{729}} & {\scriptstyle \f{743}{486}} &
{\scriptstyle \f{54608}{2187}} & 0 \\  
0 & 0 & {\scriptstyle \f{2272}{243}} & {\scriptstyle -\f{72088}{243}}
& {\scriptstyle -\f{688}{243}} & {\scriptstyle -\f{1240}{81}} &
{\scriptstyle -\f{19072}{9}} & {\scriptstyle -\f{14096}{9}} &
{\scriptstyle \f{1708}{9}} & {\scriptstyle \f{1622}{9}} & 0 &
{\scriptstyle \f{6464}{243}} & {\scriptstyle -\f{7220}{81}} &
{\scriptstyle -\f{227008}{729}} & 0 \\  
0 & 0 & {\scriptstyle \f{45424}{729}} & {\scriptstyle \f{84236}{729}}
& {\scriptstyle -\f{3880}{729}} & {\scriptstyle \f{1220}{243}} &
{\scriptstyle \f{32288}{81}} & {\scriptstyle -\f{15976}{27}} &
{\scriptstyle -\f{6692}{81}} & {\scriptstyle -\f{2437}{54}} & 0 &
{\scriptstyle \f{63824}{729}} & {\scriptstyle \f{6700}{243}} &
{\scriptstyle \f{551648}{2187}} & 0 \\  
0 & 0 & {\scriptstyle -\f{1576}{81}} & {\scriptstyle \f{446}{27}} &
{\scriptstyle \f{172}{81}} & {\scriptstyle \f{40}{27}} & 0 & 0 & 0 & 0
& {\scriptstyle \f{325}{9}} & {\scriptstyle -\f{808}{81}} &
{\scriptstyle \f{349}{27}} & {\scriptstyle -\f{8}{9}} & 0 \\  
0 & 0 & 0 & 0 & 0 & 0 & 0 & 0 & 0 & 0 & 0 & {\scriptstyle
\f{4688}{27} - 2 \beta_s^{(1)}} & 0 & 0 & 0 \\  
0 & 0 & 0 & 0 & 0 & 0 & 0 & 0 & 0 & 0 & 0 & {\scriptstyle
-\f{2192}{81}} & {\scriptstyle \f{4063}{27} - 2 \beta_s^{(1)}} & 0 & 0
\\   
0 & 0 & 0 & 0 & 0 & 0 & 0 & 0 & 0 & 0 & 0 & 0 & 0 & {\scriptstyle
-2 \beta_s^{(1)}} & 0 \\  
0 & 0 & 0 & 0 & 0 & 0 & 0 & 0 & 0 & 0 & 0 & 0 & 0 & 0 & {\scriptstyle
- 2 \beta_s^{(1)}} 
\end{array}
\right ) \, .  
\eeq
}%
The one-loop $\ord (\as)$ mixing of the electroweak penguin operators 
$Q_3^Q$--$Q_6^Q$ into $Q_9$, as well as the mixing of $Q_1^b$ has
never been computed before. The same applies to the complete two-loop
$\ord (\as^2)$ mixing\footnote{The self-mixing of $Q_3^Q$--$Q_6^Q$ has
been computed in the so-called ``standard'' operator basis in
\cite{otheradm}. Since the operator basis of the latter papers differs
from the one used here one has to perform a non-trivial
transformation. This can serve as a cross-check of the direct
calculation. See \Ref{future} for details.} of $Q_3^Q$--$Q_6^Q$ and
$Q_1^b$. As far as the remaining entries are concerned our results
agree with the well-established results in the literature
\cite{Bobeth:1999mk, Baranowski:1999tq, Gambino:2003zm,
Chetyrkin:1997gb, moreadm}.      

The matrices $\hat{\gamma}_e^{(0)}$ and $\hat{\gamma}_e^{(1)}$ are
given by 
{%
\renewcommand{\arraycolsep}{4.0pt}
\beq \label{eq:gammae0}
\hat{\gamma}_e^{(0)} = 
\left (
\begin{array}{ccccccccccccccc}
{\scriptstyle -\f{8}{3}} & 0 & 0 & 0 & 0 & 0 & {\scriptstyle
\f{32}{27}} & 0 & 0 & 0 & 0 & 0 & 0 & 0 & 0 \\  
0 & {\scriptstyle -\f{8}{3}} & 0 & 0 & 0 & 0 & {\scriptstyle
\f{8}{9}} & 0 & 0 & 0 & 0 & 0 & 0 & 0 & 0 \\  
0 & 0 & 0 & 0 & 0 & 0 & {\scriptstyle \f{76}{9}} & 0 & {\scriptstyle
-\f{2}{3}} & 0 & 0 & 0 & 0 & 0 & 0 \\  
0 & 0 & 0 & 0 & 0 & 0 & {\scriptstyle -\f{32}{27}} & {\scriptstyle
\f{20}{3}} & 0 & {\scriptstyle -\f{2}{3}} & 0 & 0 & 0 & 0 & 0 \\  
0 & 0 & 0 & 0 & 0 & 0 & {\scriptstyle \f{496}{9}} & 0 & {\scriptstyle
-\f{20}{3}} & 0 & 0 & 0 & 0 & 0 & 0 \\  
0 & 0 & 0 & 0 & 0 & 0 & {\scriptstyle -\f{512}{27}} & {\scriptstyle
\f{128}{3}} & 0 & {\scriptstyle -\f{20}{3}} & 0 & 0 & 0 & 0 & 0 \\  
0 & 0 & {\scriptstyle \f{40}{27}} & 0 & {\scriptstyle -\f{4}{27}} & 0
& {\scriptstyle \f{332}{27}} & 0 & {\scriptstyle -\f{2}{9}} & 0 & 0 &
0 & 0 & 0 & 0 \\  
0 & 0 & 0 & {\scriptstyle \f{40}{27}} & 0 & {\scriptstyle -\f{4}{27}}
& {\scriptstyle \f{32}{81}} & {\scriptstyle \f{20}{9}} & 0 &
{\scriptstyle -\f{2}{9}} & 0 & 0 & 0 & 0 & 0 \\  
0 & 0 & {\scriptstyle \f{256}{27}} & 0 & {\scriptstyle -\f{40}{27}} &
0 & {\scriptstyle \f{3152}{27}} & 0 & {\scriptstyle -\f{20}{9}} & 0 &
0 & 0 & 0 & 0 & 0 \\  
0 & 0 & 0 & {\scriptstyle \f{256}{27}} & 0 & {\scriptstyle
-\f{40}{27}} & {\scriptstyle \f{512}{81}} & {\scriptstyle \f{128}{9}}
& 0 & {\scriptstyle -\f{20}{9}} & 0 & 0 & 0 & 0 & 0 \\  
0 & 0 & 0 & 0 & 0 & 0 & {\scriptstyle -\f{16}{9}} & 0 & 0 & 0 &
{\scriptstyle \f{4}{3}} & 0 & 0 & 0 & 0 \\  
0 & 0 & 0 & 0 & 0 & 0 & 0 & 0 & 0 & 0 & 0 & {\scriptstyle \f{16}{9}} &
{\scriptstyle -\f{8}{3}} & 0 & 0 \\  
0 & 0 & 0 & 0 & 0 & 0 & 0 & 0 & 0 & 0 & 0 & 0 & {\scriptstyle
\f{8}{9}} & 0 & 0 \\  
0 & 0 & 0 & 0 & 0 & 0 & 0 & 0 & 0 & 0 & 0 & 0 & 0 & {\scriptstyle 8 -
2 \beta_e^{(0)}} & -4 \\   
0 & 0 & 0 & 0 & 0 & 0 & 0 & 0 & 0 & 0 & 0 & 0 & 0 & -4 & {\scriptstyle
-2 \beta_e^{(0)}} 
\end{array}
\right ) \, , 
\eeq
}%
{%
\renewcommand{\arraycolsep}{0.0pt}
\beq \label{eq:gammae1}
\hat{\gamma}_e^{(1)} = 
\left (
\begin{array}{ccccccccccccccc}
{\scriptstyle \f{169}{9}} & {\scriptstyle \f{100}{27}} & 0 &
{\scriptstyle \f{254}{729}} & 0 & 0 & {\scriptstyle \f{2272}{729}} &
{\scriptstyle \f{122}{81}} & 0 & {\scriptstyle \f{49}{81}} & 0 &
{\scriptstyle -\f{928}{729}} & {\scriptstyle \f{118}{243}} &
{\scriptstyle -\f{11680}{2187}} & {\scriptstyle -\f{416}{81}} \\  
{\scriptstyle \f{50}{3}} & {\scriptstyle -\f{8}{3}} & 0 &
{\scriptstyle \f{1076}{243}} & 0 & 0 & {\scriptstyle -\f{1952}{243}} &
{\scriptstyle -\f{748}{27}} & 0 & {\scriptstyle \f{82}{27}} & 0 &
{\scriptstyle -\f{232}{243}} & {\scriptstyle -\f{92}{81}} &
{\scriptstyle -\f{2920}{729}} & {\scriptstyle -\f{104}{27}} \\  
0 & 0 & 0 & {\scriptstyle \f{11116}{243}} & 0 & {\scriptstyle
-\f{14}{3}} & {\scriptstyle -\f{23488}{243}} & {\scriptstyle
\f{6280}{27}} & {\scriptstyle \f{112}{9}} & {\scriptstyle
-\f{538}{27}} & 0 & {\scriptstyle -\f{32}{243}} & {\scriptstyle
\f{32}{81}} & {\scriptstyle -\f{39752}{729}} & {\scriptstyle
-\f{136}{27}} \\  
0 & 0 & {\scriptstyle \f{280}{27}} & {\scriptstyle \f{18763}{729}} &
{\scriptstyle -\f{28}{27}} & {\scriptstyle -\f{35}{18}} &
{\scriptstyle \f{31568}{729}} & {\scriptstyle \f{9481}{81}} &
{\scriptstyle -\f{92}{27}} & {\scriptstyle -\f{1012}{81}} & 0 &
{\scriptstyle \f{64}{729}} & {\scriptstyle \f{260}{243}} &
{\scriptstyle \f{1024}{2187}} & {\scriptstyle -\f{448}{81}} \\  
0 & 0 & 0 & {\scriptstyle \f{111136}{243}} & 0 & {\scriptstyle
-\f{140}{3}} & {\scriptstyle -\f{233920}{243}} & {\scriptstyle
\f{68848}{27}} & {\scriptstyle \f{1120}{9}} & {\scriptstyle
-\f{5056}{27}} & 0 & {\scriptstyle -\f{23480}{243}} & {\scriptstyle
\f{2096}{81}} & {\scriptstyle -\f{381344}{729}} & {\scriptstyle
-\f{15616}{27}} \\  
0 & 0 & {\scriptstyle \f{2944}{27}} & {\scriptstyle \f{193312}{729}} &
{\scriptstyle -\f{280}{27}} & {\scriptstyle -\f{175}{9}} &
{\scriptstyle \f{352352}{729}} & {\scriptstyle \f{116680}{81}} &
{\scriptstyle -\f{752}{27}} & {\scriptstyle -\f{10147}{81}} & 0 &
{\scriptstyle -\f{6464}{729}} & {\scriptstyle \f{3548}{243}} &
{\scriptstyle \f{24832}{2187}} & {\scriptstyle -\f{7936}{81}} \\  
0 & 0 & {\scriptstyle -\f{2240}{81}} & {\scriptstyle \f{39392}{729}} &
{\scriptstyle \f{224}{81}} & {\scriptstyle -\f{92}{27}} &
{\scriptstyle -\f{5888}{729}} & {\scriptstyle \f{13916}{81}} &
{\scriptstyle \f{112}{27}} & {\scriptstyle -\f{812}{81}} & 0 &
{\scriptstyle -\f{544}{729}} & {\scriptstyle \f{544}{243}} &
{\scriptstyle -\f{90424}{2187}} & {\scriptstyle -\f{152}{81}} \\  
0 & 0 & {\scriptstyle \f{2176}{243}} & {\scriptstyle \f{84890}{2187}}
& {\scriptstyle -\f{184}{243}} & {\scriptstyle -\f{224}{81}} &
{\scriptstyle -\f{2552}{2187}} & {\scriptstyle \f{15638}{243}} &
{\scriptstyle -\f{176}{81}} & {\scriptstyle -\f{2881}{486}} & 0 &
{\scriptstyle -\f{64}{2187}} & {\scriptstyle -\f{260}{729}} &
{\scriptstyle -\f{1024}{6561}} & {\scriptstyle \f{448}{243}} \\  
0 & 0 & {\scriptstyle -\f{23552}{81}} & {\scriptstyle \f{399776}{729}}
& {\scriptstyle \f{2240}{81}} & {\scriptstyle -\f{752}{27}} &
{\scriptstyle -\f{90944}{729}} & {\scriptstyle \f{90128}{81}} &
{\scriptstyle \f{1120}{27}} & {\scriptstyle -\f{1748}{81}} & 0 &
{\scriptstyle -\f{28936}{729}} & {\scriptstyle \f{3664}{243}} &
{\scriptstyle -\f{910048}{2187}} & {\scriptstyle -\f{8000}{81}} \\  
0 & 0 & {\scriptstyle \f{23296}{243}} & {\scriptstyle
\f{933776}{2187}} & {\scriptstyle -\f{1504}{243}} & {\scriptstyle
-\f{2030}{81}} & {\scriptstyle \f{1312}{2187}} & {\scriptstyle
\f{102488}{243}} & {\scriptstyle -\f{1592}{81}} & {\scriptstyle
-\f{6008}{243}} & 0 & {\scriptstyle \f{6464}{2187}} & {\scriptstyle
-\f{15212}{729}} & {\scriptstyle -\f{24832}{6561}} & {\scriptstyle
\f{7936}{243}} \\  
0 & 0 & 0 & {\scriptstyle -\f{232}{81}} & 0 & 0 & 0 & {\scriptstyle
\f{580}{27}} & 0 & {\scriptstyle -\f{94}{27}} & {\scriptstyle
-\f{388}{9}} & {\scriptstyle -\f{232}{243}} & {\scriptstyle
\f{70}{81}} & {\scriptstyle \f{344}{243}} & {\scriptstyle
-\f{232}{27}} \\  
0 & 0 & 0 & 0 & 0 & 0 & 0 & 0 & 0 & 0 & 0 & {\scriptstyle
-\f{256}{27} - 2 \beta_{se}^{(1)}} & {\scriptstyle -\f{52}{9}} & 0 & 0
\\   
0 & 0 & 0 & 0 & 0 & 0 & 0 & 0 & 0 & 0 & 0 & {\scriptstyle \f{128}{81}}
& {\scriptstyle -\f{40}{27} - 2 \beta_{se}^{(1)}} & 0 & 0 \\   
0 & 0 & 0 & 0 & 0 & 0 & 0 & 0 & 0 & 0 & 0 & 0 & 0 & {\scriptstyle -2
\beta_{se}^{(1)} - 2 \beta_{es}^{(1)}} & 16 \\   
0 & 0 & 0 & 0 & 0 & 0 & 0 & 0 & 0 & 0 & 0 & 0 & 0 & 16 & {\scriptstyle
-2 \beta_{se}^{(1)} - 2 \beta_{es}^{(1)}}  
\end{array}
\right ) \, .  
\eeq
}%
In $\hat{\gamma}_e^{(0)}$ only the entries corresponding to the
self-mixing of $Q_9$ and $Q_{10}$ are new, while all other elements
agree with those of \cite{Baranowski:1999tq}. On the other hand,
$\hat{\gamma}_e^{(1)}$ is entirely new.

The two-loop ADMs $\hat \gamma^{(1)}_s$ and $\hat \gamma^{(1)}_e$
depend on the definition of evanescent operators. We have followed
the definition employed in \cite{Gambino:2003zm,Chetyrkin:1997gb} 
extending it to include all electroweak and semileptonic operators:   
\beq \label{eq:evanescent} 
\begin{split}
E_1 & = (\bar{s}_L \gamma_\mu \gamma_\nu \gamma_\rho T^a c_L)
(\bar{c}_L \gamma^\mu \gamma^\nu \gamma^\rho T^a b_L) - 16 Q_1 \, 
, \\[1mm]
E_2 & = (\bar{s}_L \gamma_\mu \gamma_\nu \gamma_\rho c_L) 
(\bar{c}_L \gamma^\mu \gamma^\nu \gamma^\rho b_L) - 16 Q_2 \, ,
\\[1mm]  
E_3 & = (\bar{s}_L \gamma_\mu \gamma_\nu \gamma_\rho \gamma_\sigma
\gamma_\tau b_L) \scalebox{1.0}{$\sum\nolimits_q$} (\bar{q} \gamma^\mu
\gamma^\nu \gamma^\rho \gamma^\sigma \gamma^\tau q) + 64 Q_3 - 20 Q_5
\, , \\[1mm] 
E_4 & = (\bar{s}_L \gamma_\mu \gamma_\nu \gamma_\rho \gamma_\sigma 
\gamma_\tau T^a b_L) \scalebox{1.0}{$\sum\nolimits_q$} (\bar{q}
\gamma^\mu \gamma^\nu \gamma^\rho \gamma^\sigma \gamma^\tau T^a q) +
64 Q_4 - 20 Q_6 \, , \\[1mm] 
E_3^Q & = (\bar{s}_L \gamma_\mu \gamma_\nu \gamma_\rho \gamma_\sigma
\gamma_\tau b_L) \scalebox{1.0}{$\sum\nolimits_q$} Q_q (\bar{q}
\gamma^\mu \gamma^\nu \gamma^\rho \gamma^\sigma \gamma^\tau q) + 64
Q_3^Q - 20 Q_5^Q \, , \\[1mm]  
E_4^Q & = (\bar{s}_L \gamma_\mu \gamma_\nu \gamma_\rho \gamma_\sigma 
\gamma_\tau T^a b_L) \scalebox{1.0}{$\sum\nolimits_q$} Q_q (\bar{q}
\gamma^\mu \gamma^\nu \gamma^\rho \gamma^\sigma \gamma^\tau T^a q) +
64 Q_4^Q - 20 Q_6^Q \, , \\[1mm] 
E_1^b & = \scalebox{1.0}{$\f{1}{12}$} (\bar{s}_L \gamma_\mu \gamma_\nu
\gamma_\rho \gamma_\sigma \gamma_\tau T^a b_L) (\bar{b} \gamma^\mu
\gamma^\nu \gamma^\rho \gamma^\sigma \gamma^\tau T^a b) +
\scalebox{1.0}{$\f{10}{3}$} (\bar{s}_L \gamma_\mu T^a b_L) (\bar{b}
\gamma^\mu T^a b) \\ 
& - \scalebox{1.0}{$\f{7}{6}$} (\bar{s}_L \gamma_\mu \gamma_\nu
\gamma_\rho T^a b_L) (\bar{b} \gamma^\mu \gamma^\nu \gamma^\rho T^a b)
- 2 Q_1^b \, , \\[1mm]
E_2^b & = (\bar{s}_L \gamma_\mu \gamma_\nu \gamma_\rho \gamma_\sigma
\gamma_\tau b_L) (\bar{b} \gamma^\mu \gamma^\nu \gamma^\rho
\gamma^\sigma \gamma^\tau b) - \scalebox{1.0}{$\f{64}{3}$} (\bar{s}_L
\gamma_\mu b_L) (\bar{b} \gamma^\mu b) \\ 
& + \scalebox{1.0}{$\f{4}{3}$} (\bar{s}_L \gamma_\mu \gamma_\nu
\gamma_\rho b_L) (\bar{b} \gamma^\mu \gamma^\nu \gamma^\rho b) - 256
Q_1^b \, , \\[1mm]  
E_9 & = \scalebox{1.0}{$\f{e^2}{\gs^2}$} (\bar{s}_L \gamma_\mu
\gamma_\nu \gamma_\rho b_L) \scalebox{1.0}{$\sum\nolimits_\ell$}
(\bar{\ell} \gamma^\mu \gamma^\nu \gamma^\rho \ell) - 10 Q_9 + 6 Q_{10}
\, , \\[1mm]  
E_{10} & = \scalebox{1.0}{$\f{e^2}{\gs^2}$} (\bar{s}_L \gamma_\mu
\gamma_\nu \gamma_\rho b_L) \scalebox{1.0}{$\sum\nolimits_\ell$}
(\bar{\ell} \gamma^\mu \gamma^\nu \gamma^\rho \gamma_5 \ell) + 6 Q_9
- 10 Q_{10} \, .  
\end{split}
\eeq

At this point a comment concerning the two-loop
mixing of $Q_1^b$ is in order. Using the definition of
$Q_1^b$, $E_1^b$ and $E_2^b$ as given in \Eqsand{list}{eq:evanescent},
the electromagnetic operator $Q_1^b$ mixes into new physical operators
beyond the one-loop level. Needless to say, such a mixing does not
alter the results presented throughout the paper. Further details on
the two-loop mixing of $Q_1^b$ will be given in a forthcoming
publication \cite{future}.

\end{document}